\documentclass[12pt]{iopart}

\usepackage{iopams}
\usepackage{setstack}
\usepackage{epsf,graphicx,rotating,color}

\newcommand{\eps}{\varepsilon}
\newcommand{\la}{\langle}
\newcommand{\lla}{\left\langle}
\newcommand{\ra}{\rangle}
\newcommand{\rra}{\right\rangle}

\newcommand{\ox}{\overset{\circ}{x}}
\newcommand{\oE}{\overset{\circ}{E}}
\newcommand{\noe}{\overset{\circ}{e}}

\begin{document}

\title[Fidelity decay]{A random matrix formulation of fidelity decay}

\author{T~Gorin\dag\ddag, T~Prosen\S{} and T~H~Seligman\ddag$\|$}

\address{\dag\ Theoretische Quantendynamik,
   Albert-Ludwigs-Universit\" at, Hermann-Herder-Str. 3, D-79104 Freiburg, 
   Germany}
\address{\ddag\ Centro Internacional de Ciencias, C.P. 62131 Cuernavaca, 
   Morelos Mexico}
\address{\S\ Physics department, Faculty of Mathematics and Physics,
   University of Ljubljana, Jadranska 19, SI-1000 Ljubljana, Slovenia }
\address{$\|$\ Centro de Ciencias Fisicas, University of Mexico (UNAM),
   C.P. 62210 Cuernavaca, Morelos, Mexico}

\ead{thomas.gorin@physik.uni-freiburg.de}

\begin{abstract}
We propose to study echo dynamics in a random matrix framework, where we 
assume that the perturbation is time independent, random and orthogonally 
invariant. This allows to use a basis in which the unperturbed Hamiltonian 
is diagonal and its properties are thus largely determined by its spectral 
statistics. We concentrate on the effect of spectral correlations usually 
associated to chaos and disregard secular variations in spectral 
density. We obtain analytic results for the fidelity decay in the linear 
response regime. To extend the domain of validity, we heuristically 
exponentiate the linear response result. The resulting expressions,
exact in the perturbative limit, are accurate approximations in the transition 
region between the ``Fermi golden rule'' and the perturbative regimes, as
examplarily verified for a deterministic chaotic system.
To sense the effect of spectral stiffness, we apply our model also to the
extreme cases of random spectra and equidistant spectra. In our analytical
approximations as well as in extensive Monte Carlo calculations, we find that 
fidelity decay is fastest for random spectra and slowest for equidistant ones, 
while the classical ensembles lie in between. We conclude that spectral
stiffness systematically enhances fidelity.
\end{abstract}

%


\maketitle

\section{\label{I} Introduction}

Loschmidt echoes, proposed more than a century ago as a gedanken 
experiment \cite{Losch1870}, have been realized experimentally~\cite{Hahn50}
measuring the corresponding auto correlation function known as  
{\it fidelity}~\cite{Peres84}. Fidelity has also been used as the simplest 
benchmark for the reliability of quantum information devices~\cite{ProZni02}, 
and recently Kaplan~\cite{Kap02} has used fidelity of eigenstates to test 
uncertainty of quantisation. Many papers study general properties of fidelity 
under different regimes, including the semiclassical~\cite{JP01,JacBee01}, and 
the linear response regime~\cite{ProZni02,CerTom02,PSZ03}. The latter can be 
divided  into the so called (standard) perturbative and the 
``Fermi golden rule'' regimes~\cite{Cohen00}. In the perturbative regime, 
linear response theory will be correct for times long as compared to the 
Heisenberg time, while in the golden rule regime, it will break down before 
the Heisenberg time. The basic aim of this paper is  to use a random 
perturbation $V$ and, following the spirit of random matrix theory, to derive 
the universal features we may expect to be relevant for chaotic systems.
The randomness of the perturbation implies orthogonal (or in the time reversal 
breaking case unitary) invariance. This in turn permits to use a basis in 
which the unperturbed Hamiltonian $H_0$ is diagonal. Then, only the spectral 
properties of $H_0$ enter the problem. We disregard secular variations in the
spectral density and concentrate on fluctuations. 

Our treatment shall make extensive use of linear response theory, because this 
will cover the basic needs of quantum information. Yet the model is not 
restricted to the lowest orders in this framework, and we shall see in Monte 
Carlo simulations how to extend heuristically our description to long time 
scales, exponentiating the linear response result. This will prove exact in
the perturbative limit and will give good results even for moderately small
perturbations. Note that in any case, we expect the linear response theory to 
hold much longer than the time corresponding to the inverse spectral span for 
which the eigenphases of the time evolution operator fill a small part of the 
unit circle, and which we call Zeno time. This regime allows for the 
straightforward expansion of the exponentials and thus is of limited interest. 
Furthermore in the context of information processing the linear 
response regime is most relevant, because in the semiclassical regime it will 
be almost impossible to have coherence in the quantum registers, and times 
short compared to the inverse spectral span are typically too short for 
quantum information processes.

We therefore construct the following model: $H_0$ is a diagonal matrix with 
an unfolded spectrum (constant level density) and $V$ is a random matrix 
pertaining to the Gaussian orthogonal or unitary ensemble (GOE or GUE). To 
model 
chaotic systems we shall follow the quantum chaos conjecture~\cite{CVG80},
and use unfolded GOE or GUE spectra for $H_0$. To contrast these with the 
possible extreme cases, we shall also consider picket-fence (equally spaced) 
and random spectra. Actually, any spectrum can be used, whose form factor is 
known. Random spectra are often associated with integrable 
systems~\cite{BerTab77}, but we must keep in mind that the effects we describe 
in echo-dynamics are state dependent, and thus other important properties of 
integrable systems such as invariant tori will typically also be relevant. We 
can therefore expect the GOE and the GUE cases to relate directly to chaotic 
systems, and indeed we shall see that such a relation can be established, as 
shown in~\cite{CerTom03}. Indeed, the random matrix model used there, is very
similar to the one we propose. Note that the use of a random perturbation has
been proposed in reference~\cite{ProZni02}. 

The other important aspect in echo dynamics is the dependence on the initial 
state. The analysis of coherent states -- so important in other studies -- 
is not considered here, because, typically, we do not expect to have an 
underlying classical model at our disposal. The most important property is 
then the 
spectral span of the initial state. We shall mainly consider two extremes: the 
evolution of an eigenstate of the unperturbed system, and the evolution of a 
random state with a given, but fairly large spectral span. The latter is the 
relevant case for quantum computing. This is easily understood if we think 
{\it e.g.} of Fourier transforms. The transforms of simple functions such as
Gaussians are readily done analytically, while we would like to exploit the 
quantum algorithm for an arbitrarily complicated case. Yet it  
is also interesting to consider eigenfunctions partially because of the
applications in the context of effects of different types of 
quantisation \cite{Kap02} but also because these display most 
markedly the effects of spectral fluctuations.  

We measure the distortion of a state by calculating the fidelity, which 
may be interpreted as the autocorrelation function of a forward evolution with 
$H_0$ and a time reversed  evolution with the perturbed Hamiltonian 
$H=H_0 +\lambda V$. The same quantity can also be interpreted as a cross 
correlation function of the same initial state evolving under the two 
Hamiltonians $H_0$ and $H$. Specifically, if $\Psi(t)$ and $\Psi_0(t)$ are the 
functions evolving under $H$ and $H_0$, respectively, from the same initial 
function $\Psi(0) = \Psi_0(0)$, we define the {\it fidelity amplitude} as the 
matrix element
\begin{equation}
f(t) = \la\Psi_0(t)|\Psi(t)\ra  = \la\Psi(0)| U_0(-t)\, U(t)|\Psi(0)\ra \; , 
\label{defH} \end{equation}
where $U_0(t)$ is the unitary propagator associated with $H_0$, and $U(t)$
is the one associated with $H$. From the point of view of a quantum Loschmidt 
echo, $U_0(-t)\, U(t)$ would be the ``echo''-operator. The fidelity 
is usually defined as $F(t)= |f(t)|^2$. In the interaction picture with the 
state of the system 
denoted by $x(t)$, the fidelity is simply the autocorrelation function. To see 
this we recall that 
\begin{equation}
\Psi(t) = U_0(t) \; x(t) \;\Rightarrow\;
\rmi\hbar\partial_t \; x(t) = \lambda \tilde V(t) \; x(t) \; ,
\end{equation}
where $\tilde V(t) = U_0(-t)\; V\; U_0(t)$. Hence:
\begin{equation}
f(t) = \la x(0)|U_0(-t)|\Psi(t)\ra = \la x(0)|x(t)\ra \; .
\label{defF}\end{equation}
In the next section we recall the relevant linear response relations in 
the interaction picture, and apply them to the calculation of the fidelity 
amplitude. In the perturbative regime, {\it i.e.} for small $\lambda$, the 
Born series can be summed to yield the well known Gaussian decay. We therefore 
give a phenomenological formula for the general case by simply exponentiating 
the linear response result, and we compare to Monte Carlo simulations to 
establish the accuracy
of this formula. In section~\ref{CT}, we apply our results to the 
standard map, using calculations by Cerruti and Tomsovic~\cite{CerTom03}.
We find that our heuristic formula is not only simpler, but also
closer to the numerical experiment than the expression
derived there.

In section~\ref{F}, we proceed to calculate the fidelity. We observe
a marked difference between the behaviour of a single eigenstate and that of a 
superposition. In the former case, the differences resulting from different 
spectral statistics are enhanced.
Again we compare with numerical calculations. 
In section~\ref{T} we study the asymptotic behaviour of the fidelity for 
long times in various situations, and we obtain some analytical results from 
the perturbation theory. All these studies use spectral correlation 
functions in the large $N$ limit for the GOE (and GUE) case. 
Finally we give some concluding remarks in section~\ref{C}.

\section{\label{A} Fidelity amplitude}

In this section we shall first derive a linear response formula for the
decay of the fidelity amplitude. The central quantity of interest is here
the correlation integral, which is a double integral over the autocorrelation
function of the perturbation in the interaction picture. Given that the
perturbation is taken from the GOE (GUE), the correlation integral only
depends on the spectral properties of $H_0$, and we shall evaluate the 
correlation integral for different cases. Note that, unless stated otherwise,
GOE perturbations are used throughout this article. We then discuss the 
possibility to 
extend the validity of the linear response result by packing it into an 
exponential. This phenomenological formula works very well up to the crossover 
between the perturbative and the Fermi golden rule regime. In the last part of 
this section we discuss the sample fluctuations of the fidelity amplitude 
concentrating on the regime of high fidelity, as this is the one most relevant
for quantum information applications.

As a preliminary step, let us introduce more convenient units for time
and energy. In order to take advantage of the orthogonal (unitary) 
invariance of the perturbation, we write the Schr\" odinger equation in
the eigenbasis of $H_0$:
\begin{equation}
H_0 = {\rm diag}(\oE_\alpha) \; , \quad U_0(t) = 
{\rm diag}(\rme^{-\rmi \oE_\alpha t/\hbar}) \; .
\end{equation}
Thus, we obtain in the interaction picture:
\begin{equation}
\rmi\hbar\,\partial_t\; x(t)= \lambda\; 
   {\rm diag}(\rme^{\rmi\oE_\alpha\, t/\hbar})\; V\;
   {\rm diag}(\rme^{-\rmi\oE_\beta\, t/\hbar})\; x(t) \; .
\end{equation}
Denoting with $d$ the
average level spacing in the spectrum of $H_0$ (for simplicity we shall assume
that $d$ is constant in the energy range of interest), the Heisenberg time is
defined as $t_H = 2\pi\hbar/d$. It is convenient to measure time in units of 
$t_H$ and energy in units of $d$, introducing the dimensionless time $s= t/t_H$
and spectrum $\{\noe_\alpha = \oE_\alpha/d\}$. Then we obtain
for $x'(s)= x(t_H\, s)$:
\begin{equation}
\rmi\,\partial_s\; x'(s) = \frac{2\pi\lambda}{d}\; 
   {\rm diag}(\rme^{2\pi\, \rmi\noe_\alpha\, s})\; V\;
   {\rm diag}(\rme^{-2\pi\rmi\, \noe_\beta\, s})\; x'(s) \; .
\end{equation}
This shows, that without restriction of generality, we may assume that the
Heisenberg time and the average level spacing are both equal to one -- and
in the rest of the paper we shall indeed do so. This allows us to stick to
the symbols for time and energy, as introduced in the beginning.

\subsection{\label{AL} Linear response}

With $U_0(t)= {\rm diag}[\exp(-2\pi\rmi\, \oE_\alpha\, t)]$, and the
shorthand $\tilde V(t)= U_0(-t)\, V\, U_0(t)$, we can approximate $x(t)$ 
using the Born series:
\begin{equation}
x^{(n)}(t) = x(0) -2\pi\rmi\, \lambda\int_0^t\rmd\tau\; \tilde V(\tau)\; 
x^{(n-1)}(\tau) \; ,\quad x^{(0)}(t) = x(0) \; .
\end{equation}
Up to second order we obtain: $x^{(2)}(t) = X(t)\; x(0)$, where
\begin{equation}
X(t) = 1 -2\pi\rmi\, \lambda\int_0^t\rmd\tau \; \tilde V(\tau) - 4\pi^2\,
\lambda^2
\int_0^t\rmd\tau\int_0^\tau\rmd\tau'\; \tilde V(\tau)\, \tilde V(\tau') \; .
\label{defX}\end{equation}
Note that $X(t)$ is the linear response approximation of the echo operator, 
mentioned above. In general, we are interested in the various moments of the 
components of $x^{(2)}(t)$. Their knowledge will allow us to calculate 
averages of the fidelity, as well as the survival probability or even the 
purity. However, we may additionally choose the initial state $x(0)$ to be 
random and eventually even $H_0$ could be a member of some ensemble. In 
particular, if we want to keep the initial states arbitrary, it is more 
convenient to consider the various moments of the matrix elements 
$X_{\alpha\beta}(t)$, when the average is performed over the GOE (or GUE) 
matrix $V$. We define these matrix ensembles such that 
\begin{equation}
\la V_{ij}\, V_{kl}\ra = 
   \cases{\delta_{il}\delta_{jk}                           &: GUE\\
          \delta_{ik}\delta_{jl} +  \delta_{il}\delta_{jk} &: GOE}  \; .
\label{AL_GEdef}\end{equation}
Our linear response approach is very similar to the one by Prosen and 
coworkers 
\cite{ProZni01}. However, in their work, $H_0$ and $V$ are fixed while the 
average is taken over initial conditions only. Here we will primarily 
average over the perturbation $V$, thereafter over $H_0$, if we consider an
ensemble, and finally over initial states, if these are random.

\subsubsection*{The spectral correlation function}

In order to average the linear response operator $X(t)$, given in (\ref{defX}),
we basically need the correlation 
function: $\la\tilde V(\tau)\,\tilde V(\tau')\ra_V$, where $\la\ldots\ra_V$ 
denotes the ensemble average over the random matrix $V$. Here, we present 
the full calculation for the GOE perturbation, only. 
With $\Delta_\alpha(t)= \exp(-2\pi\rmi\, \oE_\alpha\, t)$, we obtain:
\begin{eqnarray}
\fl \la [\tilde V(\tau)\,\tilde V(\tau')]_{\alpha\beta}\ra_V
 = \sum_\gamma \Delta_\alpha(-\tau) \; \la V_{\alpha\gamma} \; 
   \Delta_\gamma(\tau-\tau')\; V_{\gamma\beta}\ra_V \; \Delta_\beta(\tau') 
   \nonumber\\
\lo= \sum_\gamma \Delta_\alpha(-\tau) \; 
   \Delta_\gamma(\tau-\tau') \;
   \Delta_\beta(\tau') \; [\delta_{\alpha\gamma}\delta_{\gamma\beta}
 +  \delta_{\alpha\beta}] \nonumber\\
\lo= \delta_{\alpha\beta} \; C_\alpha(\tau'-\tau) \; , \quad
C_\alpha(t) = 1 + \sum_\gamma \rme^{2\pi\rmi(\oE_\gamma-\oE_\alpha)\, t} \; . 
\label{calC}\end{eqnarray}
The correlation function~(\ref{calC}) can be 
expressed in terms of the two-point form factor $b_2(t)$, one of the important 
fluctuation measures in quantum chaos studies \cite{Meh91}. Namely, if we 
denote with $\la\ldots\ra_0$ the average over the diagonal matrix $H_0$, we 
may write:
\begin{equation}
\la C_\alpha(t)\ra_0 = 2 + \delta(t) - b_2(t) \; .
\label{defb2}\end{equation}
The $H_0$-average makes sure to obtain a well 
behaved, {\it i.e.} smooth, two-point form factor. Though, note that in
practice, this additional averaging may be avoidable. The corresponding 
expression for the GUE perturbation is: $1 + \delta(t) - b_2(t)$.

\nosections

\noindent
Let us now turn to the calculation of the average echo operator in the linear
response approximation:
\begin{equation}
\la X_{\alpha\beta}(t)\ra_V = \delta_{\alpha\beta} - 4\pi^2\, \lambda^2
   \int_0^t\rmd\tau\int_0^\tau\rmd\tau'\; \delta_{\alpha\beta}\; 
   C_\alpha(\tau') \; .
\end{equation}
Additional averaging over $H_0$ gives a scalar result:
\begin{equation}
\la\la X_{\alpha\beta}(t)\ra_V\ra_0 = 1 - 4\pi^2\, \lambda^2\; 
{\cal C}(t)\; , 
\label{AL:chole}\end{equation}
where
\begin{equation}
{\cal C}(t)= t^2+ t/2 - \int_0^t\rmd\tau\int_0^\tau\rmd\tau'\; b_2(\tau') \; .
\label{cint}\end{equation}
Note that the integration of the delta function gives $\case{1}{2}$. 
For the GUE perturbation, the $t^2$-term should be cut in half.
Due to the scalar result in equation~(\ref{AL:chole}), the average fidelity 
amplitude does not depend on the initial state:
\begin{equation}
\la\la f(t)\ra_V\ra_0 = \la\la\la X(t)\ra_V\ra_0\ra_{\ox} 
 = 1 - 4\pi^2\, \lambda^2\; {\cal C}(t) \; .
\label{A:lresp}\end{equation}
Here $\la\ldots\ra_{\ox}$ is a short hand for the expectation value with 
respect to the initial state.
If we avoid averaging over $H_0$, the average fidelity amplitude reads:
\begin{equation}
\la f(t)\ra_V = \la \la X(t)\ra_{\ox} \ra_V = 1- 4\pi^2\, \lambda^2 
   \int_0^t\rmd\tau\int_0^\tau\rmd\tau' \;
   \sum_\alpha |\ox_\alpha|^2\,  C_\alpha(\tau') \; .
\end{equation}
If the participation ratio of the initial state $\ox$ is 
sufficiently large, {\it i.e.} if the inverse participation ratio, 
${\rm ipr}(\ox)= \sum_\alpha |\ox_\alpha|^4$, is sufficiently small, this may
lead to self-averaging -- just as it does for the autocorrelation 
function of certain deterministic systems. Then we may still use the
general result, equation~(\ref{A:lresp}), inserting the proper two-point form 
factor for this case.


\subsection{Correlation integral}

\begin{figure}
\begin{center}
\setlength{\unitlength}{1pt}
\begin{picture}(445,157)
\put(20,20){\includegraphics[scale=0.5]{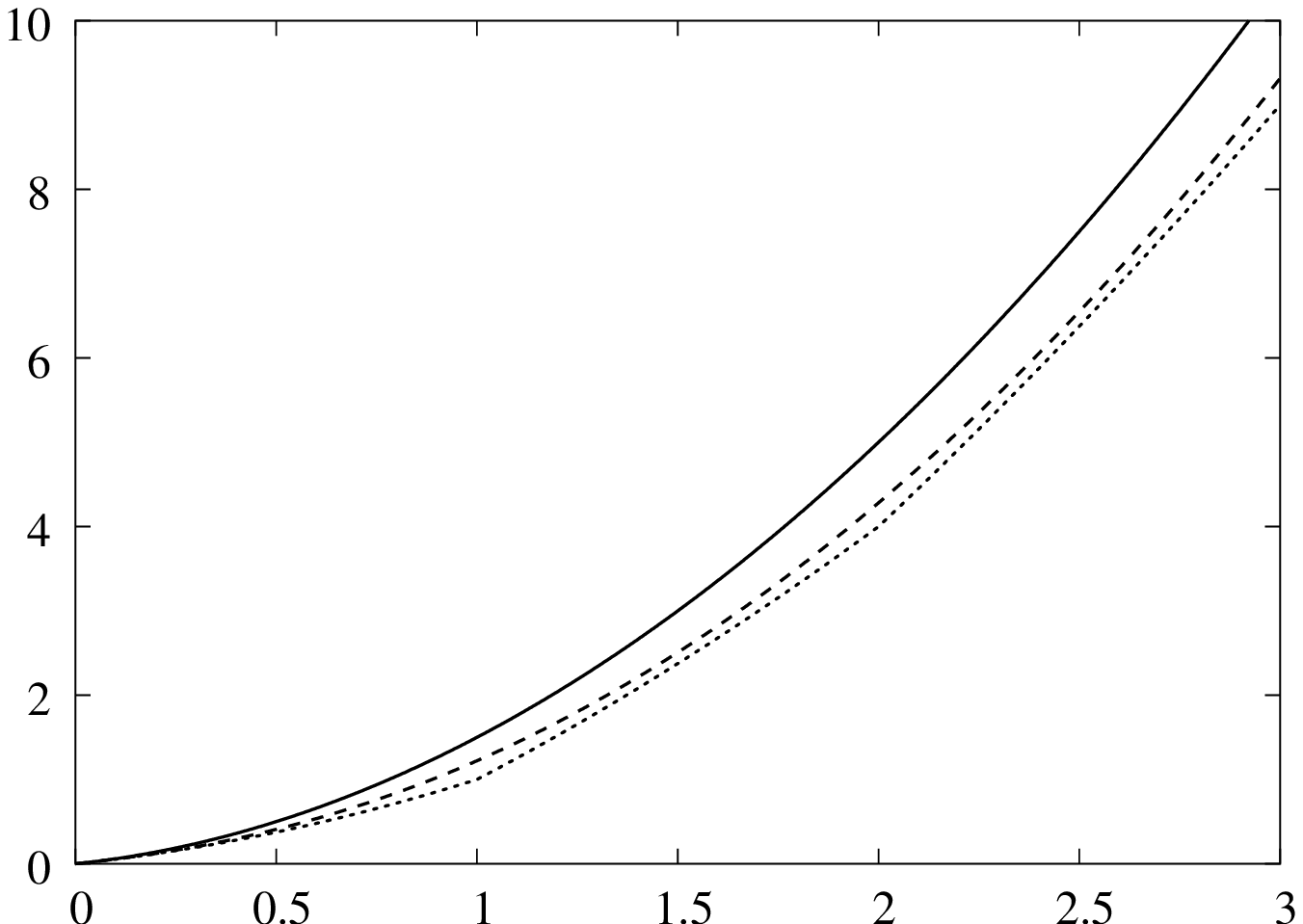}}
\put(244,20){\includegraphics[scale=0.5]{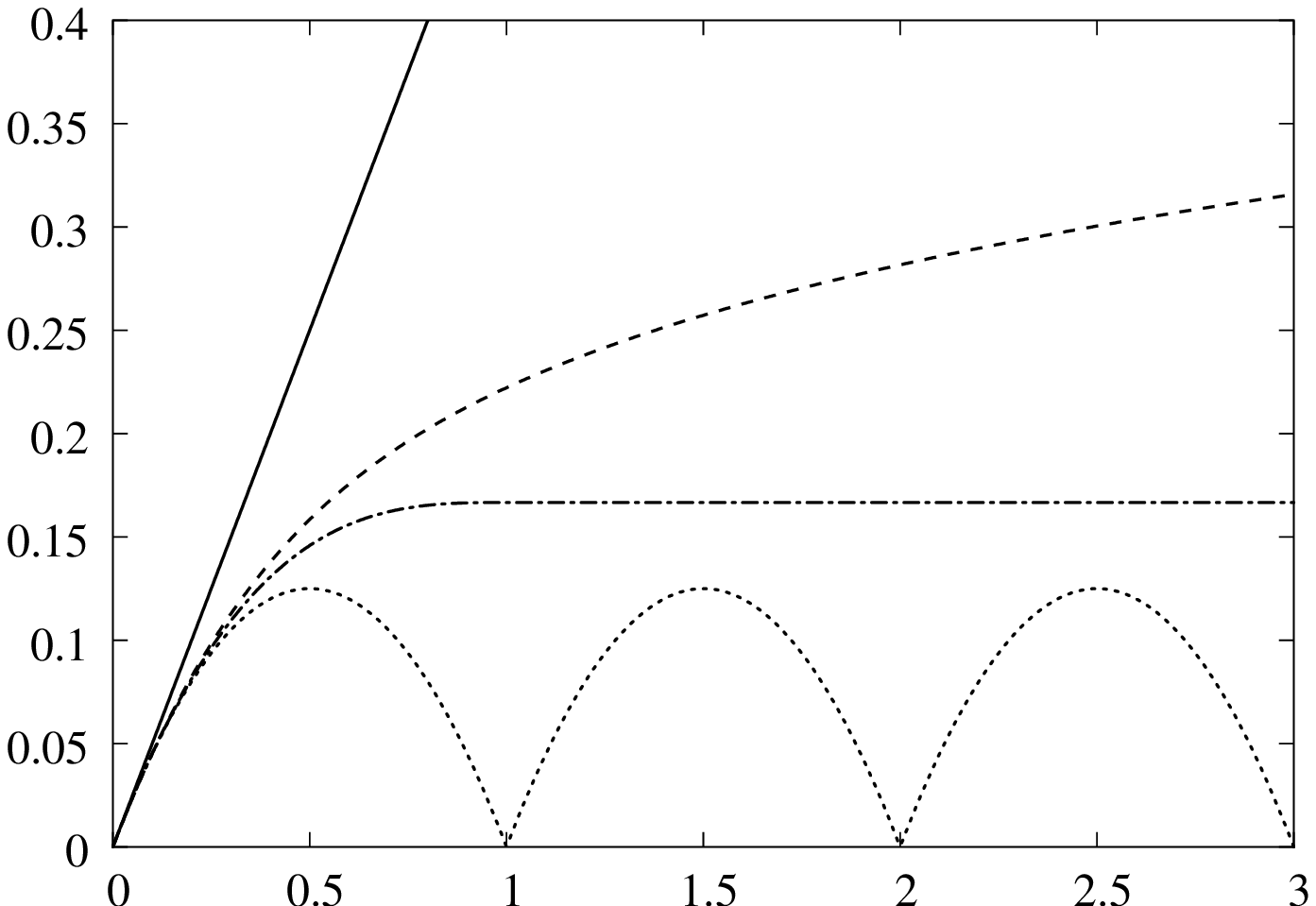}}
\put(187,145){\makebox(0,0){(a)}}
\put(420,145){\makebox(0,0){(b)}}
\put(128,10){\makebox(0,0){$t$}}
\put(353,10){\makebox(0,0){$t$}}
\put(10,90){\makebox(0,0){\begin{sideways} ${\cal C}(t)$\end{sideways}}}
\put(234,90){\makebox(0,0){\begin{sideways} 
                              ${\cal C}(t) - t^2$\end{sideways}}}
\end{picture}
\caption{\label{corrint-1} (a) The correlation integral ${\cal C}(t)$, 
equation~(\ref{cint}), for the Poisson spectrum (solid line), the GOE spectrum 
(dashed line), and the picket-fence spectrum (dotted line). (b) The function 
${\cal C}(t) - t^2$, for the three spectra in panel (a), and 
for the GUE spectrum (dash-dotted line).}  
\end{center}
\end{figure}

Clearly, the correlation integral ${\cal C}(t)$ is the essential quantity
which determines the fidelity (amplitude) decay. It is shown in 
figure~\ref{corrint-1} for the Poisson, GOE, GUE, and picket-fence spectrum. 
The two-point form factor $b_2(t)$ is zero in the Poisson
case, while for all other cases, it is given in~\ref{A_Cint}. The level 
repulsion, present in all but the first spectrum, tends to slow down 
the increase of the correlation integral. This is simply a consequence of the 
so called ``correlation hole''~\cite{LLJP86}, which can be found in the 
integrand of equation~(\ref{AL:chole}). Note however that for large times 
the quadratic increase with time remains essentially unaffected [panel~(a)]. 
In figure~\ref{corrint-1}(b), we subtract the $t^2$-term in order
to display more clearly the particular effects of the different spectral
correlations. While in the Poisson case the 
remaining term is linear, in the correlated spectra (GOE, GUE and picket-fence)
the linear increase at the origin is capped. In fact, we can calculate
explicitely the asymptotic behaviour of the correlation integral. In the GOE
case, using the exact formula~(\ref{AGOE:BB}), we obtain:
\begin{equation}
{\cal C}(t) = t^2 + \frac{t}{2} - \int_0^t\rmd\tau\; B(\tau) 
   = t^2 + \frac{\ln t + 2 + \ln 2}{12} + \Or(t^{-1}) \; .
\end{equation}
In the picket-fence case, the correlation integral (with the $t^2$ term
subtracted) has zeros at integer values of $t$. The GUE spectrum leads to 
${\cal C}(t) - t^2$ approaching $2\pi^2/3$ from below.

\begin{figure}
\begin{center}
\setlength{\unitlength}{1pt}
\begin{picture}(260,185)
\put(20,20){\includegraphics[scale=0.6]{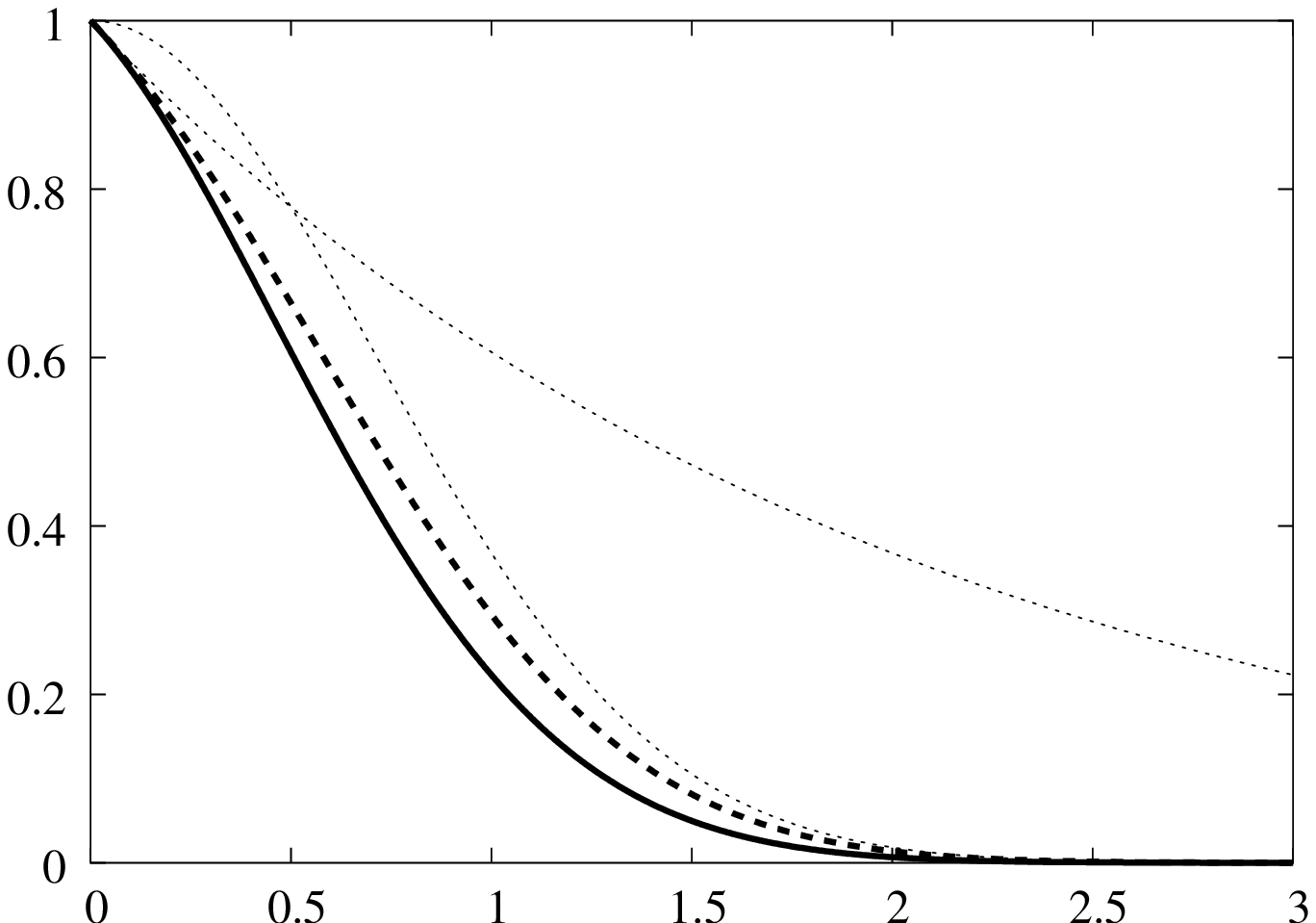}}
\put(115,87){\includegraphics[scale=0.32]{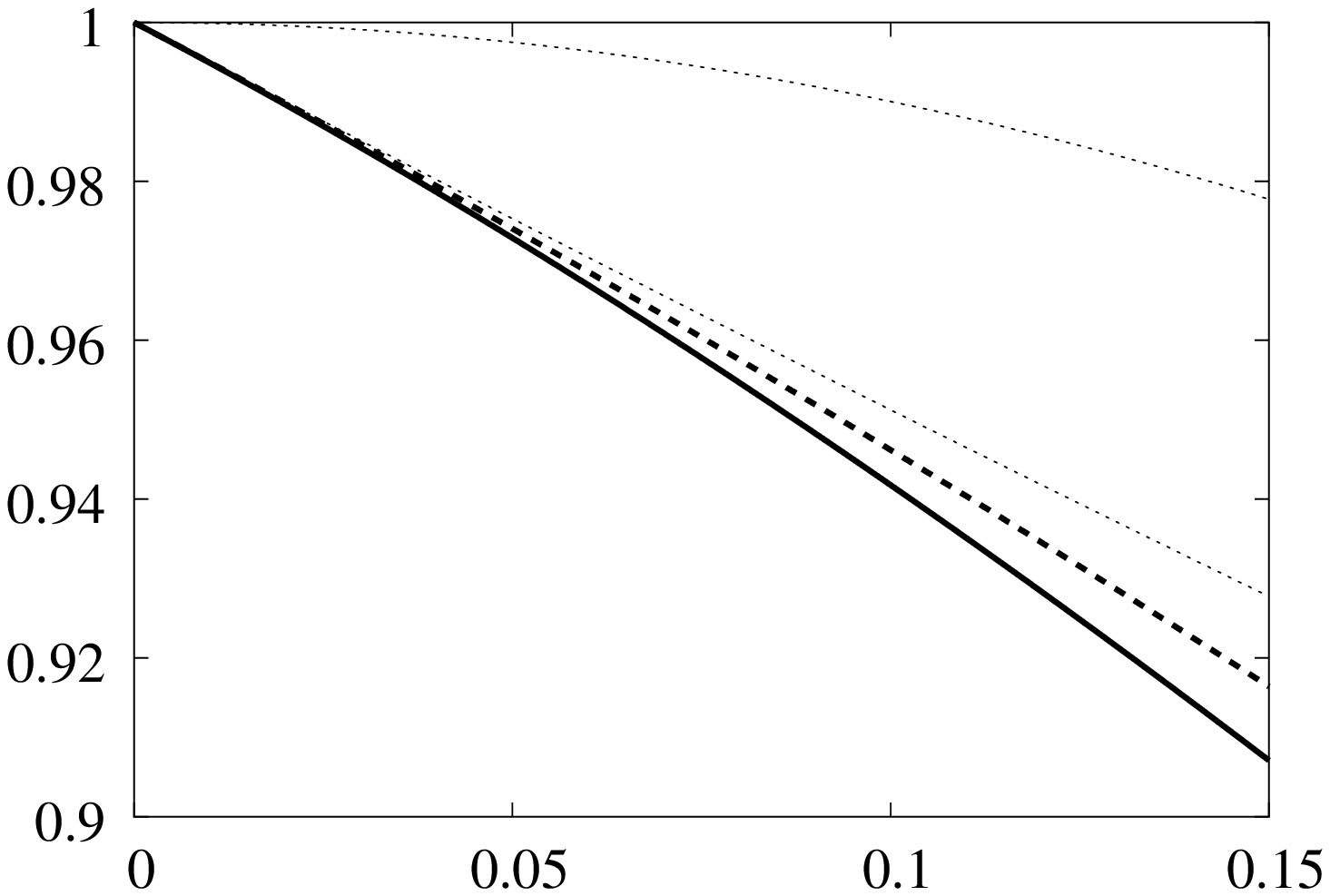}}
\put(148,10){\makebox(0,0){$t$}}
\put(10,100){\makebox(0,0){\begin{sideways} $\la f(t)\ra$\end{sideways}}}
\end{picture}
\caption{\label{crossover} The fidelity amplitude in the crossover regime, 
from linear to quadratic decay. $2\pi\lambda = 1$. Thick solid and dashed 
lines: Theory for the Poisson and the GOE case respectively, {\it i.e.}
equation~(\ref{A:lresp}) packed into an exponential (see text for details).
Thin dotted lines: Theory for the Poisson case, where either the linear term 
or the quadratic term has been ignored. The inset shows a zoom into the area
around $t=0,\, \la f(t)\ra = 1$.}
\end{center}
\end{figure}

\subsection{Exponentiated linear response}

In the perturbative regime, {\it i.e.} in the limit of small 
$\lambda$, the fidelity decay is known to be 
Gaussian~\cite{Peres84}. This can readily be reproduced in the
present context. The Born series can be trivially summed under the
assumption that eigenstates are not affected by the perturbation. The result
corresponds to the exponentiation of equation~(\ref{A:lresp}):
\begin{equation}
\la f(t)\ra = \exp[-\, 4\pi^2\, \lambda^2\, {\cal C}(t)] \; .
\label{A:eresp}\end{equation}
This result is exact in the limit $2\pi\lambda\ll 1$, only, but we shall show 
below that errors are fairly small even for $\lambda\sim 0.1$, and negligible
for $\lambda\sim 0.01$. 
The essential point is that the correlation integral, whether 
used in equation~(\ref{A:lresp}) or (\ref{A:eresp}) clearly displays the 
transition between a linear decay at short times and a quadratic decay at 
long times. Accordingly, for small $\lambda$, the decay of the fidelity 
amplitude is dominantly Gaussian, whereas for larger $\lambda$, it is 
dominantly exponential~\cite{JacBee01,CerTom02}.

Figure~\ref{crossover} shows the behaviour of the fidelity amplitude according
to equation~(\ref{A:eresp}) for the GOE and the Poisson case, in the large
$N$ limit. For the latter, the linear and quadratic terms are also shown
separately. Note that the GOE results lie systematically above the Poisson 
ones due to the correlations in the spectrum of $H_0$. The crossover 
region is precisely the one of most interest and, therefore, we check the 
accuracy of the phenomenological expression~(\ref{A:eresp}) beyond the 
perturbative limit. 
\begin{figure}
\begin{center}
\setlength{\unitlength}{1pt}
\begin{picture}(445,157)
\put(17,20){\includegraphics[scale=0.5]{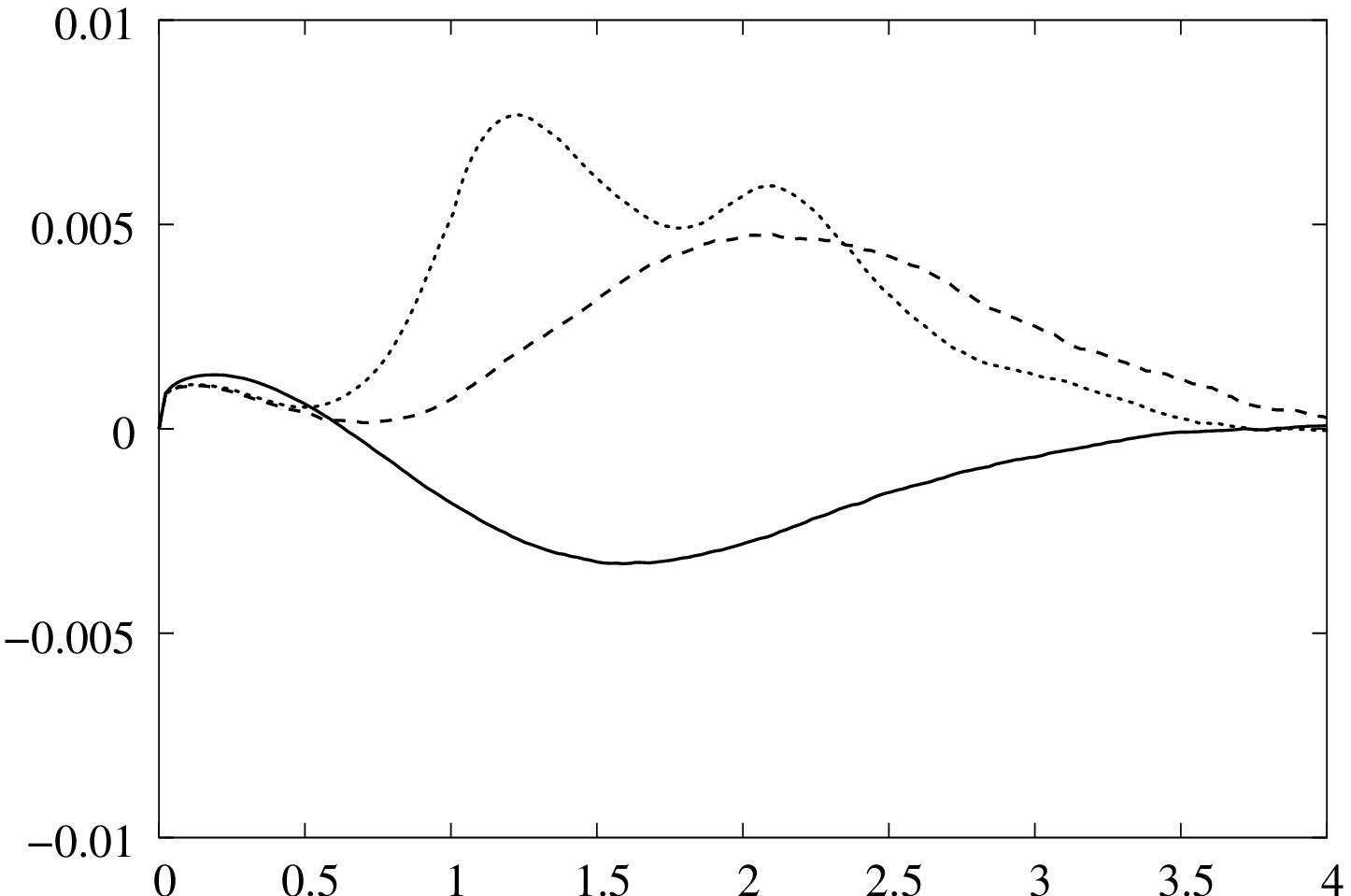}}
\put(230,20){\includegraphics[scale=0.5]{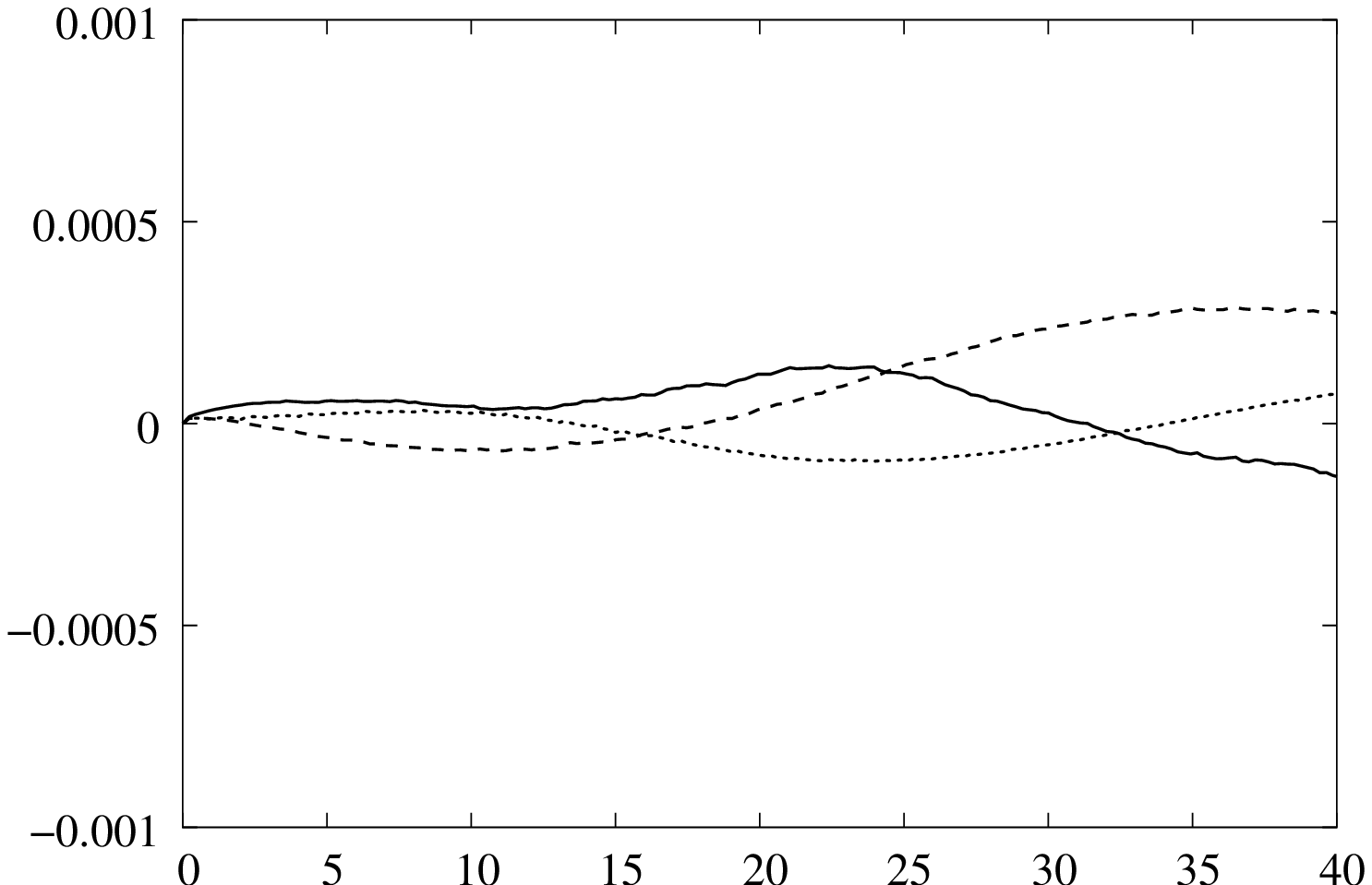}}
\put(200,145){\makebox(0,0){(a)}}
\put(420,145){\makebox(0,0){(b)}}
\put(128,10){\makebox(0,0){$t$}}
\put(353,10){\makebox(0,0){$t$}}
\put(6,90){\makebox(0,0){\begin{sideways} 
                             $\la f(t)\ra - \la f(t)\ra_{\rm theo}$
                          \end{sideways}}}
\end{picture}
\caption{\label{f:err} Difference between the Monte Carlo and the 
theoretical result, equation~(\ref{A:eresp}), for the fidelity amplitude, for 
the Poisson (solid line), the GOE (dashed line) and the picket-fence (dotted 
line) case. In (a) $\lambda = 0.1$, whereas in (b) $\lambda = 0.01$
(note the different scale on the ordinate).}
\end{center}
\end{figure}

Figure~\ref{f:err} shows the 
deviation of the fidelity amplitude obtained for Monte Carlo calculations for 
$N=100$ from what we find using equation~(\ref{A:eresp}) and the $N=\infty$ 
correlation integral. Besides the GOE case, we considered also spectra without
correlations (Poisson case) and a spectrum with equally spaced levels 
(picket-fence case). In this way we cover a broad range of possible 
spectral correlations. 
For $\lambda = 0.1$, figure~\ref{f:err}(a), the deviations are of the order
$0.5\%$ of the maximal value for the fidelity amplitude. We checked that these
deviations are not due to finite size effects. By varying $N$ and also by
considering the correlation integrals for finite $N$ (for the Poisson
and the picket-fence case), we found that those effects do not alter the error
noticeably, except for a small ``hump'' at $t\approx 0.25$ in panel (a). In 
panel (b), the error is shown for $\lambda = 0.01$. Its absolute value is 
further reduced by a factor of ten, roughly, and finite size effects are no
longer observable. Note that the errors for the random and the picket-fence 
spectrum are of the same order of magnitude as for the chaotic case. This 
suggests, that our exponentiated formula works well in all 
cases, where the two-point correlations are known. In order to compare the 
error curves in figure~\ref{f:err} with the actual value of the fidelity
amplitude, note that the time $t_{\rm half}$ where the fidelity has dropped to 
half of its initial value is about $t_{\rm half} \approx 1.1$ in panel (a),
and  $t_{\rm half} \approx 13$ in panel (b). In all, we see that the results 
shown in figure~\ref{crossover} are very similar to what we can expect from a 
Monte Carlo calculation.

\subsection{Sample fluctuations}

In the figures~\ref{corrint-1}~and~\ref{crossover} we saw the effects of 
spectral correlations on the decay of the average fidelity. But are these
effects observable also for an individual system? To answer this question, we
shall now study the fluctuations of the fidelity amplitude, focusing on
the regime of high fidelity.
 
To obtain the numerical results shown in figure~\ref{f:err} and hereafter, we 
average over the GOE perturbation $V$, the 
spectrum of $H_0$ and eventually over initial states $\ox$. The initial state 
may be an eigenstate of $H_0$; then we choose it from the centre of the 
spectrum in order to minimise border effects. Alternatively, it may 
be a random (orthogonally invariant) state. Though the average value of the 
fidelity amplitude does not depend on the choice of the initial state, higher
moments (such as the variance) do.

\begin{figure}
\begin{center}
\setlength{\unitlength}{1pt}
\begin{picture}(445,157)
\put(18,20){\includegraphics[scale=0.5]{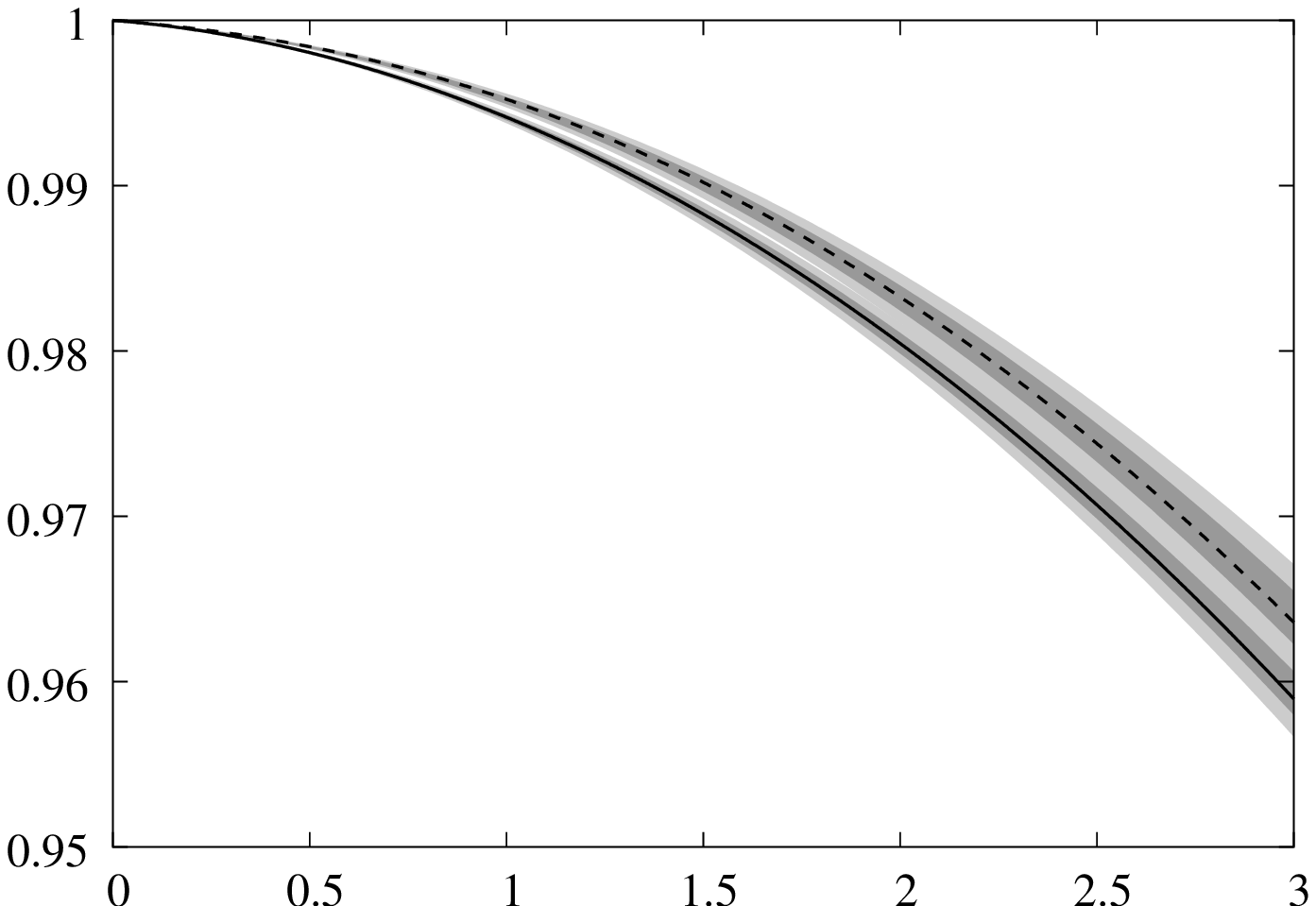}}
\put(236,20){\includegraphics[scale=0.5]{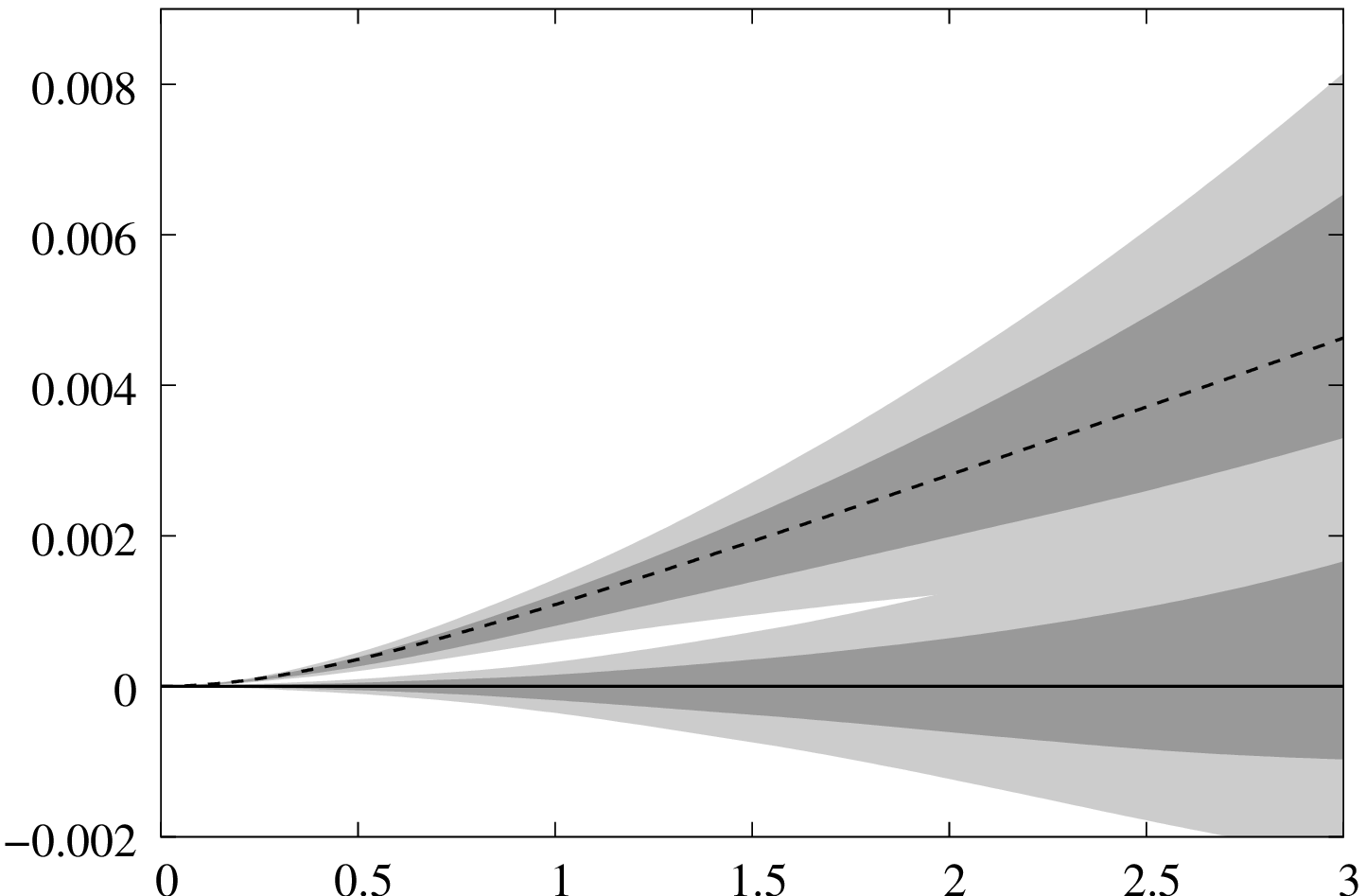}}
\put(200,145){\makebox(0,0){(a)}}
\put(410,145){\makebox(0,0){(b)}}
\put(128,10){\makebox(0,0){$t$}}
\put(353,10){\makebox(0,0){$t$}}
\put(8,90){\makebox(0,0){\begin{sideways} $\la f(t)\ra$\end{sideways}}}
\put(229,90){\makebox(0,0){\begin{sideways} 
                          $\la f(t)\ra - \la f(t)\ra_{\rm Poi}$\end{sideways}}}
\end{picture}
\caption{\label{f_A1}(a) Fidelity amplitude for $N=100, \lambda=0.01$; the 
initial states are eigenstates of $H_0$. The thick solid and dashed lines show 
the theoretical result (\ref{A:lresp}) for the Poisson and the 
GOE case, respectively. The numerical results (shaded areas) 
are obtained from $10$ small sample averages of size $n_{\rm run}= 1000$, each.
The total average of all samples lies in the centre of the gray bands, while 
their borders are given by plus/minus one standard deviation (dark gray), and 
plus/minus two standard deviations (light gray).  (b) The same quantities as 
in (a), but with the Poisson theory subtracted.}
\end{center}
\end{figure}

Figure~\ref{f_A1} shows the average fidelity amplitude $\la f(t)\ra$ in the
perturbative regime for the Poisson and the GOE case. We performed $10$ 
independent ensemble averages, each of them over $n_{\rm run}= 1000$
random systems. From this we calculated the statistical uncertainty for
a single ensemble average. In this figure, as well as in similar ones below,
the area which is not more than one (two) standard deviations away from the
total average over all samples, is plotted in dark (light) gray. In addition,
we plotted the pure linear response result, equation~(\ref{A:lresp}) for the
Poisson ensemble (solid line) and the GOE (dashed line). While in panel~(a), 
we show $f(t)$ itself, in panel~(b), the Poisson theory is subtracted in order
to show more clearly the differences between the Poisson and the GOE case on
the one hand, as well as between numerics and linear response theory on the
other.

The large sample-to-sample fluctuations are clearly due to the special choice
of the initial state. In linear response theory, and after the 
averaging over $V$, it can be seen that the fidelity amplitude depends on only 
$N$ out of $N(N-1)$ available eigenvalue differences [see 
equation~(\ref{calC})]. By consequence, we need very large samples in order to
obtain accurate averages. However, the statistical fluctuations can also be 
reduced by probing the decay of $f(t)$ at the same system for various 
initial $H_0$-eigenstates, or by using a random initial state, as shown below.
 
\begin{figure}
\begin{center}
\setlength{\unitlength}{1pt}
\begin{picture}(418,157)
\put(20,20){\includegraphics[scale=0.5]{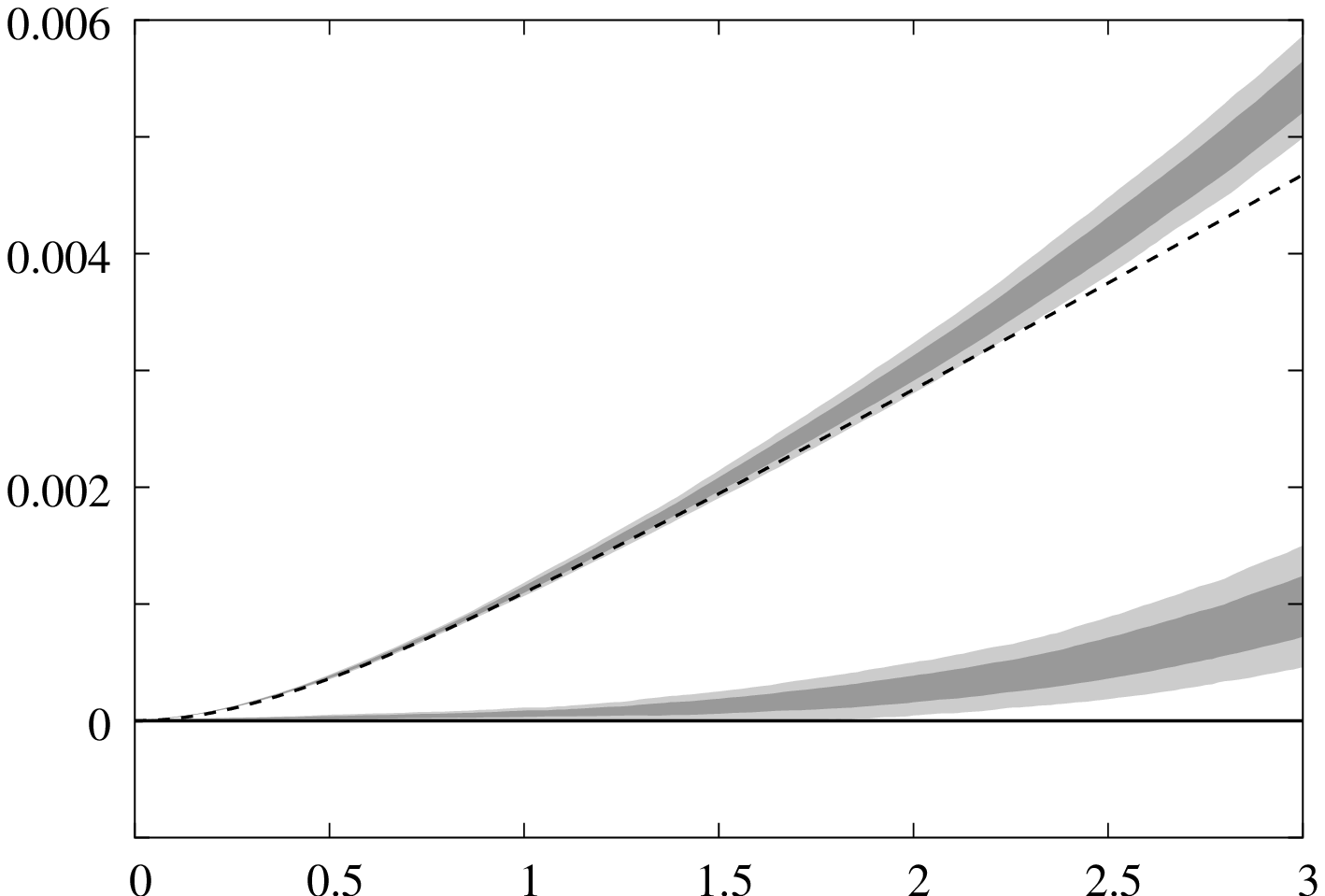}}
\put(230,20){\includegraphics[scale=0.5]{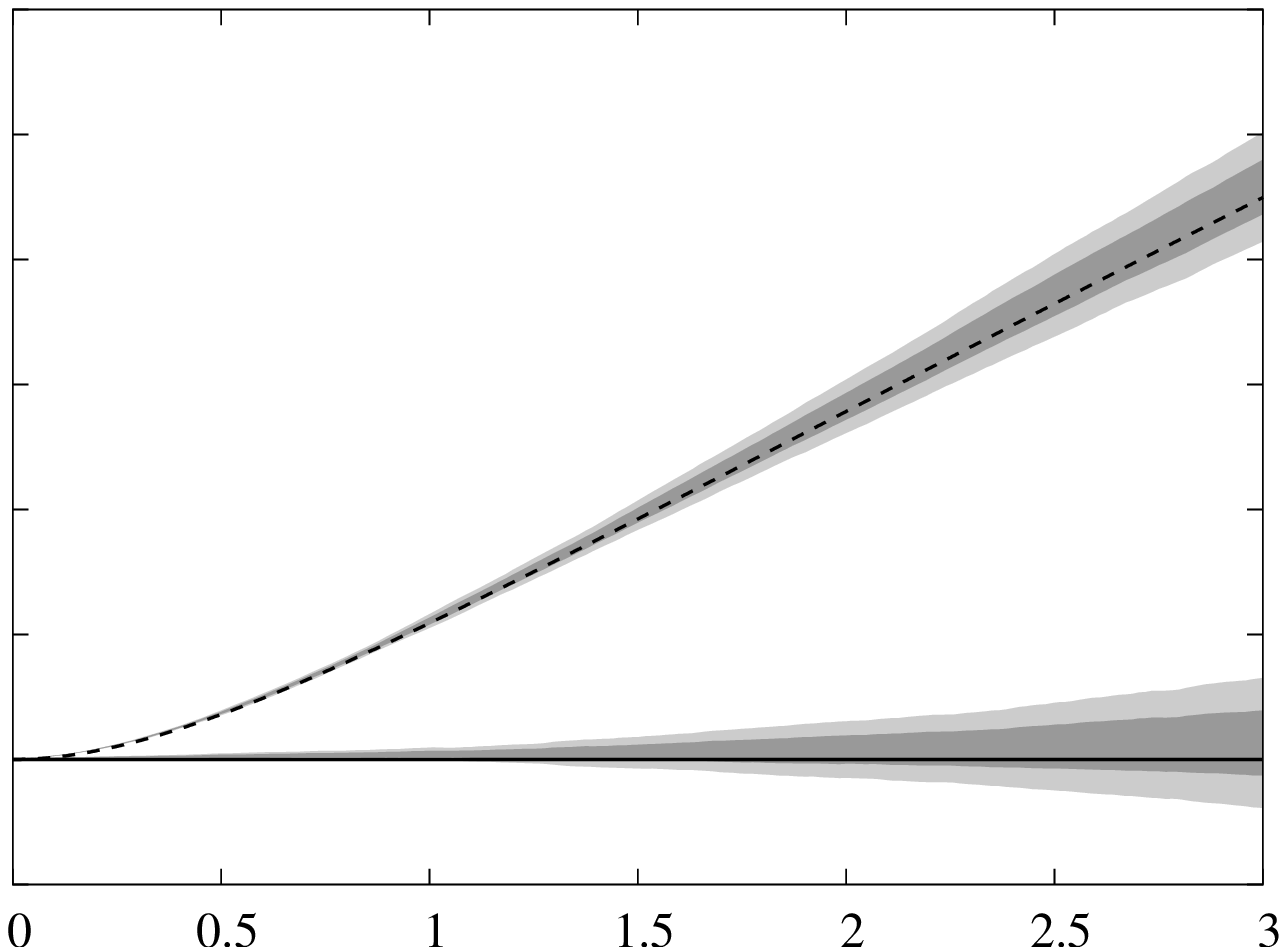}}
\put(190,145){\makebox(0,0){(a)}}
\put(390,145){\makebox(0,0){(b)}}
\put(133,10){\makebox(0,0){$t$}}
\put(324,10){\makebox(0,0){$t$}}
\put(10,90){\makebox(0,0){\begin{sideways} 
                         $\la f(t)\ra - \la f(t)\ra_{\rm Poi}$\end{sideways}}}
\end{picture}
\caption{\label{f_A2} Fidelity amplitude with the Poisson theory subtracted, 
for $N=100, \lambda= 0.01$; random initial states. The
numerical results (shaded areas) are obtained from $10$ small samples of size 
$n_{\rm run}= 1000$, each. The significance of the gray scales is the same
as in figure~\ref{f_A1}. (a) The thick solid and dashed lines show the 
theoretical results for the Poisson case (here, we expect zero) and the GOE 
case, based on the linear response result~(\ref{A:lresp}). (b) The same data, 
but with the linear response result being packed into an 
exponential~(\ref{A:eresp}).}
\end{center}
\end{figure}

Figure~\ref{f_A2} shows the fidelity amplitude (with the Poisson theory 
subtracted), for the Poisson and the GOE case, where the initial 
states are taken to be random. We used the same sample size as in 
figure~\ref{f_A1}. Nevertheless, the statistical fluctuations have drastically
decreased, and the Poisson- and the GOE-data can now be distinguished without 
difficulty. While in panel~(a) we used the pure linear response
theory, equation~(\ref{A:lresp}), for the Poisson and the GOE case, in
panel~(b), we used the exponentiated version, equation~(\ref{A:eresp}). One
can clearly see that there are systematic differences between the numerical
data and pure linear response theory, while we obtained a near perfect
agreement with the help of the exponentiation.

Using a result from the following section~\ref{F}, one can understand the 
reduction of the statistical fluctuations. 
According to equation~(\ref{F:lresp}), the variance of the 
fidelity amplitude is proportional to the IPR of the initial state. As for a 
random state $\ox\, : {\rm ipr}(\ox) = 3/(N+2)$, it implies that the 
statistical deviations should be about $\sqrt{(N+2)/3} \approx 6$ times 
smaller than in figure~\ref{f_A1}(b). We finally note that the error which
remains after exponentiating equation~(\ref{A:lresp}) is so small, that it
will typically be lost in these fluctuations.

%

\begin{figure}
\begin{center}
\setlength{\unitlength}{1pt}
\begin{picture}(260,157)
\put(20,20){\includegraphics[scale=0.5]{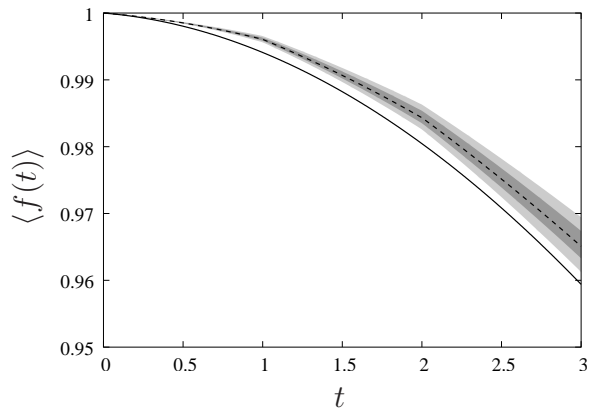}}
\put(128,10){\makebox(0,0){$t$}}
\put(10,90){\makebox(0,0){\begin{sideways} $\la f(t)\ra$\end{sideways}}}
\end{picture}
\caption{\label{f_A3pif} Fidelity amplitude for $N=100, \lambda=0.01$; small 
samples of size $n_{\rm run}= 1000$. The numerical results are represented by
shaded bands, where the significance of the gray scales is the same as in 
figure~\ref{f_A1}. Data for the picket-fence spectra, where the initial 
states are eigenstates of $H_0$. The linear response result~(\ref{A:lresp}) 
for the Poisson case (thick solid line) and the picket-fence case (thick 
dashed line).}
\end{center}
\end{figure}
%

From figure~\ref{f_A3pif} we can see that, also for the picket-fence spectra, 
the linear response theory does a good job. What is more interesting, is a 
comparison of the statistical deviations of the numerical curves for this 
figure, and figure~\ref{f_A1}(a) which is practically the same plot, but for 
the Poisson and the GOE spectra. As the spectrum in the picket-fence 
case is no statistical quantity, it means that the statistical deviations 
are mainly due to the fluctuations in the perturbation matrix $V$. This 
implies that the fidelity decay is sensitive to the level correlations in a 
single spectrum (of very moderate length, $N=100$). One would just have to 
probe this particular system with different initial states, and different 
perturbations. Quantum dots with variable magnetic fields
may be an appropriate test ground.

\nosections

\noindent
From the numerical analysis of the sample fluctuations, as well as from the 
linear response result for the variance of the fidelity amplitude, we 
conclude that the sample fluctuations are proportional to the inverse 
participation ratio of the initial state $\ox$. To observe a significant 
difference between a GOE spectrum and a spectrum without correlations, we 
need either very large samples (for initial $H_0$-eigenstates), or initial 
states with very small inverse participation ratio ({\it e.g.} random 
initial states). The sample fluctuations are mainly due to the randomness 
of the perturbation $V$, not so much due to the random $H_0$. Finally, we
have verified that the exponentiation of the linear response result improves 
the agreement with numerics considerably. In section~\ref{CT} to follow,
we will perform a further, more realistic, test.

\subsection{\label{CT} A deterministic example}

In reference~\cite{CerTom03} Cerruti and Tomsovic calculate the fidelity 
amplitude for the quantised standard map. They also present an analytical 
formula for the decay of the fidelity amplitude, which is based on a 
``uniform approximation''. In essence, they use a random matrix model similar
to ours, and an additional semiclassical argument which allows them to 
determine the perturbation strength. Their final formula agrees with ours in 
both limit cases of weak and strong perturbation, if we disregard the spectral 
correlations, {\it i.e.} if we set the two-point form factor in the 
correlation integral, equation~(\ref{cint}), equal to zero. However, in the 
crossover region, there are further deviations.

In order to compare our approach to theirs, we first have to adapt our model
to a two-fold symmetry in the standard map, which leads to two independent
symmetry sectors in the Hamiltonian. This influences the correlation
integral, and by re-examining equation~(\ref{calC}), we obtain:
\begin{equation}
{\cal C}_{\rm sym}(t)=  t^2 + \frac{1}{4}\left[ t -\int_0^t
   \rmd\tau\int_0^\tau\rmd\tau'\; b_2^{\rm GOE}(\tau') \right] \; ,
\end{equation}
where the unit for time is again chosen such that $t_H = 1$. Finally, we have
to take into account that our strength parameter $\lambda$, is related to 
Cerruti and Tomsovic's by: $\lambda^2 = 4\, \Lambda$, and that in 
reference~\cite{CerTom03} the unit for time is chosen differently.
In all, we obtain for the fidelity amplitude:
\begin{equation}
f_{\rm sym}(t) = \exp\!\!\left[- 16\, \pi^2\, \Lambda\; 
   {\cal C}_{\rm sym}(t)\right] \; ,
\label{CT-eq}\end{equation}
where Cerruti and Tomsovic's semiclassical argument gives 
$\Lambda \approx 0.0395$. The same result follows also from general
considerations of ~\cite{ProZni02}.

\begin{figure}
\begin{center}
\setlength{\unitlength}{1pt}
\begin{picture}(445,157)
\put(17,20){\includegraphics[scale=0.5]{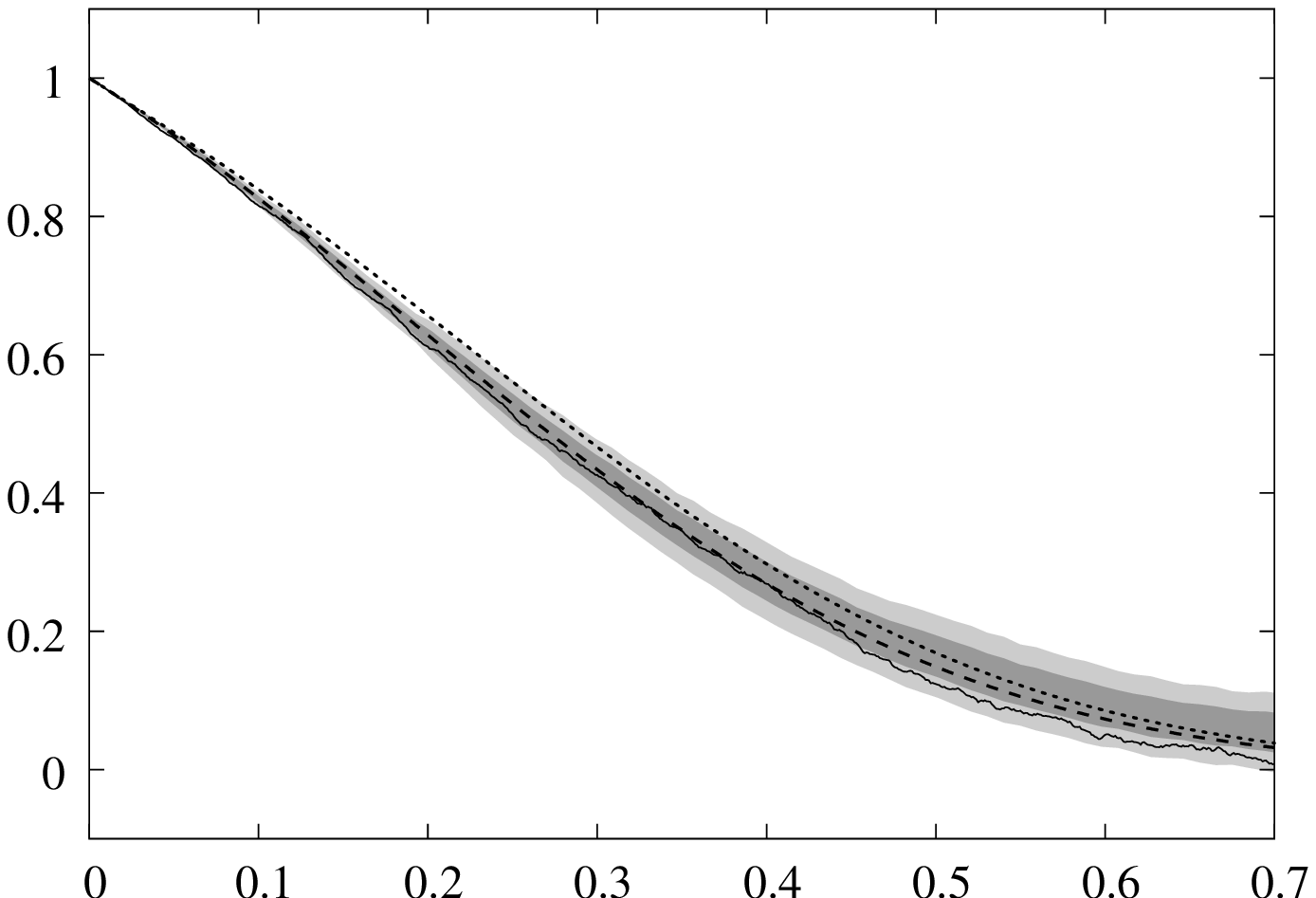}}
\put(240,20){\includegraphics[scale=0.5]{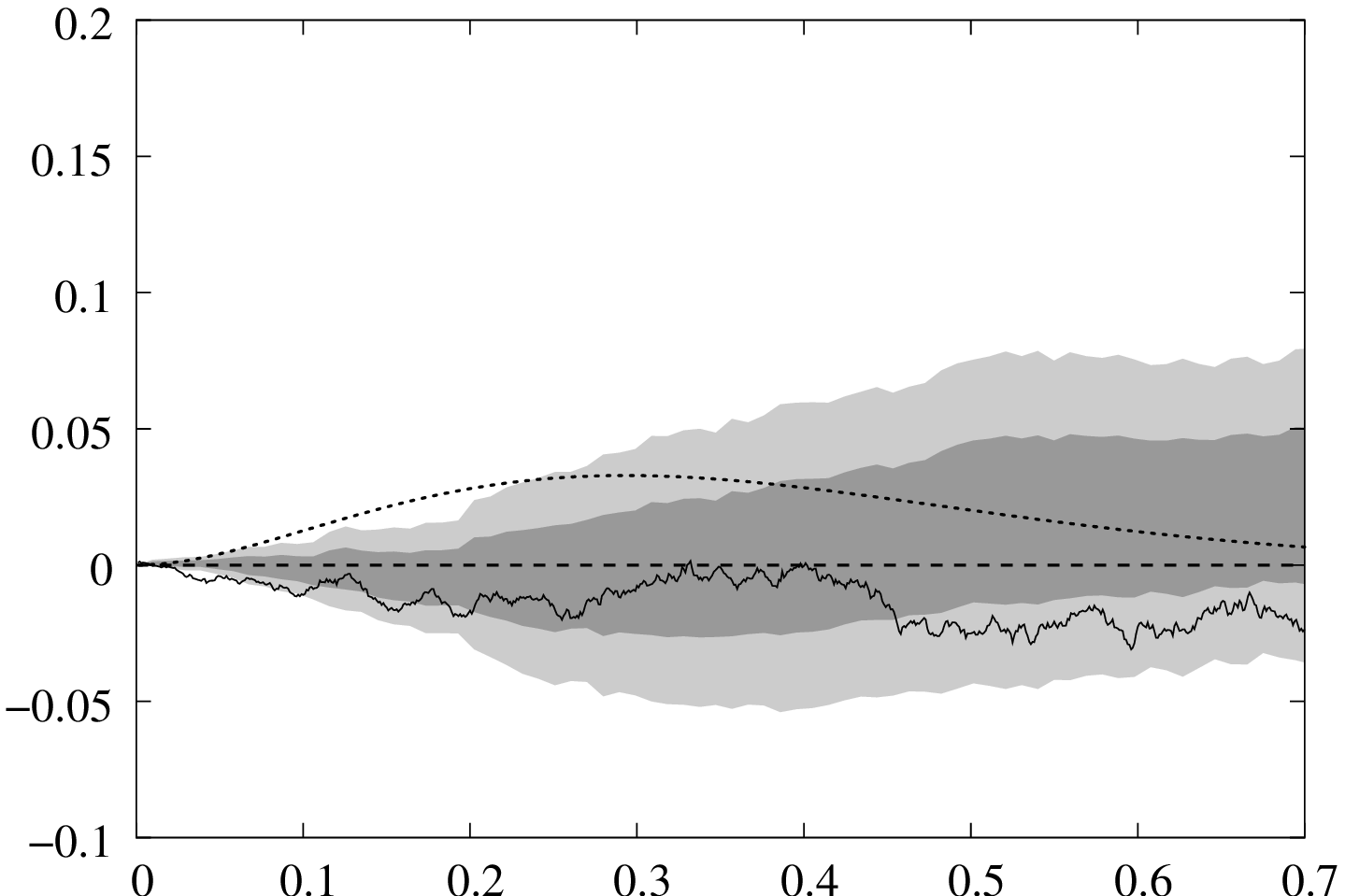}}
\put(200,145){\makebox(0,0){(a)}}
\put(420,145){\makebox(0,0){(b)}}
\put(128,10){\makebox(0,0){$t$}}
\put(353,10){\makebox(0,0){$t$}}
\put(6,90){\makebox(0,0){\begin{sideways} $f(t)$ \end{sideways}}} 
\put(230,90){\makebox(0,0){\begin{sideways} 
                             $f(t) - f_{\rm sym}(t)$
                          \end{sideways}}}
\end{picture}
\caption{\label{f:CT} (a) The fidelity amplitude for the quantised standard
map (data from~\cite{CerTom03}) (solid line). Our exponentiated linear 
response result, equation~(\ref{CT-eq}) for $\Lambda = 0.0395$ (dashed line), 
and the theoretical result of Cerruti and Tomsovic (dotted line). The average 
over the full sample lies in the centre of the gray bands, while 
their borders are given by plus/minus one standard deviation (dark gray), and 
plus/minus two standard deviations (light gray).
(b) The same curves as in (a) but with our theory, equation~(\ref{CT-eq}),
subtracted.}
\end{center}
\end{figure}

In figure~\ref{f:CT} we reproduce the numerical data for the quantised
standard map from Cerruti and Tomsovic, and compare it with a random matrix
calculation and our theory, equation~(\ref{CT-eq}). In the random matrix model
we choose a matrix of dimension $N=1000$, which is block-diagonal. Each block
contains a matrix of dimension $N/2$, produced from an unfolded GOE spectrum
(with average level spacing equal to two) and a GOE perturbation, as defined 
earlier. The variance of the non-zero off-diagonal matrix elements is 
$\lambda^2 = 4\,\Lambda$. To estimate the statistical error on the fidelity
amplitude, we calculated $f(t)$ for ten different random matrices of this
type. In addition, we plot the corrected theoretical 
result from Cerruti and Tomsovic~\cite{CerTom03}.

The fidelity decay for the quantised standard map is well described by our
random matrix model. In particular in panel~(b) we can see that the 
corresponding curve stays well within the statistical limits for individual
members of the random matrix ensemble. Though our theory is not exact (cf.
figure~\ref{f:err}, and note that $\lambda\approx 0.4$ is quite large),
the systematic deviations from the random matrix results, as well as from the
quantised standard map calculation, are very small. The theory of Cerruti and
Tomsovic shows somewhat larger systematic deviations, in particular in the
region around $t=0.2$. Note that we compare with the semiclassical result of
reference~\cite{CerTom03}, because we use the same classical parameters in
our random matrix model.

\section{\label{F} The fidelity}

In an experiment it is usually easier to measure the fidelity 
$F(t) = |f(t)|^2$ rather than the fidelity amplitude. 
Though, often it is assumed that the average fidelity is simply given by the
absolute value squared of the average fidelity amplitude, and we shall see
under which conditions this is justified.
In distinction to the average amplitude, the average fidelity depends on the 
choice of the initial state. For instance, if the initial state is 
an eigenstate of $H_0$, the fidelity $F(t)$ coincides with the survival 
probability or autocorrelation function.

\subsection{Linear response}

In order to calculate the fidelity $F(t)$, we first expand the linear
response approximation of the echo operator~(\ref{defX}) according to 
increasing powers of $\lambda$:
\begin{eqnarray}
X(t)= 1 - 2\pi\rmi\, \lambda \; I(t) - 4\pi^2\, \lambda^2\; J(t) \; ,
\label{Xdecomp}\\
I(t)= \int_0^t\rmd\tau\; \tilde V(\tau) \qquad
J(t)= \int_0^t\rmd\tau\int_0^\tau\rmd\tau' \; \tilde V(\tau)\, \tilde V(\tau') 
\; .
\label{AL:J}\end{eqnarray}
For an arbitrary initial state $\ox$, the average fidelity reads:
\begin{equation}
\fl \la F(t)\ra_V = \la\; \la X(t)\ra_{\ox}\; \la X(t)\ra_{\ox}^*\ra_V
 = 1 - 4\pi^2\, \lambda^2\left( 2{\rm Re}\, \la\la J(t)\ra_{\ox}\ra_V - 
       \la |\la I(t)\ra_{\ox}|^2\ra_V \right) \; ,
\end{equation}
where we considered only terms up to second order. For 
the average $\la\la J(t)\ra_{\ox}\ra_V$ we obtain (cf. section~\ref{AL}):
\begin{equation}
\fl \la J_{\alpha\beta}(t)\ra_V = \delta_{\alpha\beta}
   \int_0^t\rmd\tau\int_0^\tau\rmd\tau' \; C_\alpha(\tau-\tau')
= \delta_{\alpha\beta}\int_0^t\rmd\tau\int_0^\tau\rmd\tau' \; C_\alpha(\tau') 
\; ,
\end{equation}
while
\begin{equation}
\la |\la I(t)\ra_{\ox}|^2\ra_V = \sum_{\alpha\beta\gamma\eps}
   \ox_\alpha^*\, \ox_\beta\, \ox_\gamma\, \ox_\eps^* \int_0^t\rmd\tau
   \int_0^t\rmd\tau'\;
   \la\tilde V_{\alpha\beta}(\tau)\tilde V_{\gamma\eps}^*(\tau') \ra_V \; .
\end{equation}
The average product of two matrix elements of $\tilde V$ can be computed with
the help of equation~(\ref{AL_GEdef}). For the GOE perturbation, we obtain:
\begin{eqnarray}
\fl \la \tilde V_{\alpha\beta}(\tau)\,\tilde V_{\gamma\eps}^*(\tau')\ra_V
= \left[ \delta_{\alpha\gamma}\,\delta_{\beta\eps} \;
   \Delta_\alpha(-\tau+\tau')\;\Delta_\beta(\tau-\tau') +
   \delta_{\alpha\eps}\,\delta_{\beta\gamma}\; \Delta_\alpha(-\tau-\tau') \;
   \Delta_\beta(\tau+\tau')\right] \nonumber\\
\lo= \left[ \delta_{\alpha\gamma}\,\delta_{\beta\eps} \;
   \rme^{2\pi\rmi(\oE_\alpha-\oE_\beta)(\tau-\tau')} +
   \delta_{\alpha\eps}\,\delta_{\beta\gamma}\;
   \rme^{2\pi\rmi(\oE_\alpha-\oE_\beta)(\tau+\tau')} \right] \nonumber\\
\lo= \cases{                            2 &: $\alpha =\beta = \gamma =\eps$ \\
           \rme^{2\pi\rmi(\oE_\alpha-\oE_\beta)\, (\tau-\tau')} &: 
                                           $\alpha =\gamma \ne \beta =\eps$ \\
           \rme^{2\pi\rmi(\oE_\alpha-\oE_\beta)\, (\tau+\tau')} &: 
                                           $\alpha =\eps \ne \beta =\gamma$ \\
                                       0 &: otherwise} \; ,
\label{eq26}\end{eqnarray}
which leads to
\begin{eqnarray}
\fl \la |\la I(t)\ra_{\ox}|^2\ra_V = \int_0^t\rmd\tau\int_0^t\rmd\tau' 
   \left[ 2\sum_\alpha |\ox_\alpha|^4 \right. \nonumber\\
+ \left. \sum_{\alpha\ne\beta} \left( |\ox_\alpha|^2\; 
      |\ox_\beta|^2\; \rme^{2\pi\rmi(\oE_\alpha-\oE_\beta)\, (\tau-\tau')} + 
   (\ox_\alpha^*)^2\; \ox_\beta^2\; 
      \rme^{2\pi\rmi(\oE_\alpha-\oE_\beta)\, (\tau+\tau')} \right) \right] \; .
\label{eq27}\end{eqnarray}
Here $\sum_\alpha |\ox_\alpha|^4 = {\rm ipr}(\ox)$ is the inverse 
participation ratio (IPR) of the initial state $\ox$. If the initial state is 
an eigenstate of $H_0$, then ${\rm ipr}(\ox) = 1$, and this is the only term 
which survives. If $\ox$ is complex, then we expect that term which 
depends on $(\tau+\tau')$ to vanish. On the average, the contribution of 
each of the exponential 
terms is only of order $N^{-1}$. Therefore, the whole expression in the 
second line of equation~(\ref{eq27}) is of order $N^{-1}$. In the limit of
large $N$, it can be neglected, leading to
\begin{eqnarray}
\la F(t)\ra_V &= 1 - 2 (2\pi\lambda)^2\left\{ 
   {\cal C}(t) - {\rm ipr}(\ox) \; t^2 \right\} \nonumber\\
&= |\la f(t)\ra_V|^2 + 2(2\pi\lambda)^2\; t^2\; {\rm ipr}(\ox) + \Or(t^4) \; .
\label{F:lresp}\end{eqnarray}
For the GUE perturbation, equation~(\ref{eq26}) would simply read:
$\la\tilde V_{\alpha\beta}(\tau)\,\tilde V_{\gamma\eps}^*(\tau')\ra_V = 
\delta_{\alpha\beta}\, \exp[2\pi\rmi (\oE_\alpha-\oE_\beta)\, (\tau-\tau')]$.
For the average fidelity, it would have the only consequence that all terms 
involving ${\rm ipr}(\ox)$ had to be multiplied by one half.

The second line of equation~(\ref{F:lresp}) 
gives the variance of the sample fluctuations of the fidelity 
amplitude, discussed in the previous section. It shows that, if the
IPR of the initial state goes to zero, then indeed, the average fidelity is 
given by the absolute value squared of the fidelity amplitude (at least in
the linear response regime). By contrast, if the IPR of the initial state
is large (few principal components), then the behaviour of the fidelity is
quite different from that of the fidelity amplitude. In the extreme case
that the IPR is equal to one (initial state is an eigenstate of $H_0$),
the IPR-term in equation~(\ref{F:lresp}) completely kills the $t^2$-term in
the correlation integral ${\cal C}(t)$. This holds equally true whether $V$ 
is chosen from the GOE or the GUE. 

\subsection{Average fidelity and sample-to-sample fluctuations}

\begin{figure}
\begin{center}
\setlength{\unitlength}{1pt}
\begin{picture}(418,157)
\put(20,20){\includegraphics[scale=0.5]{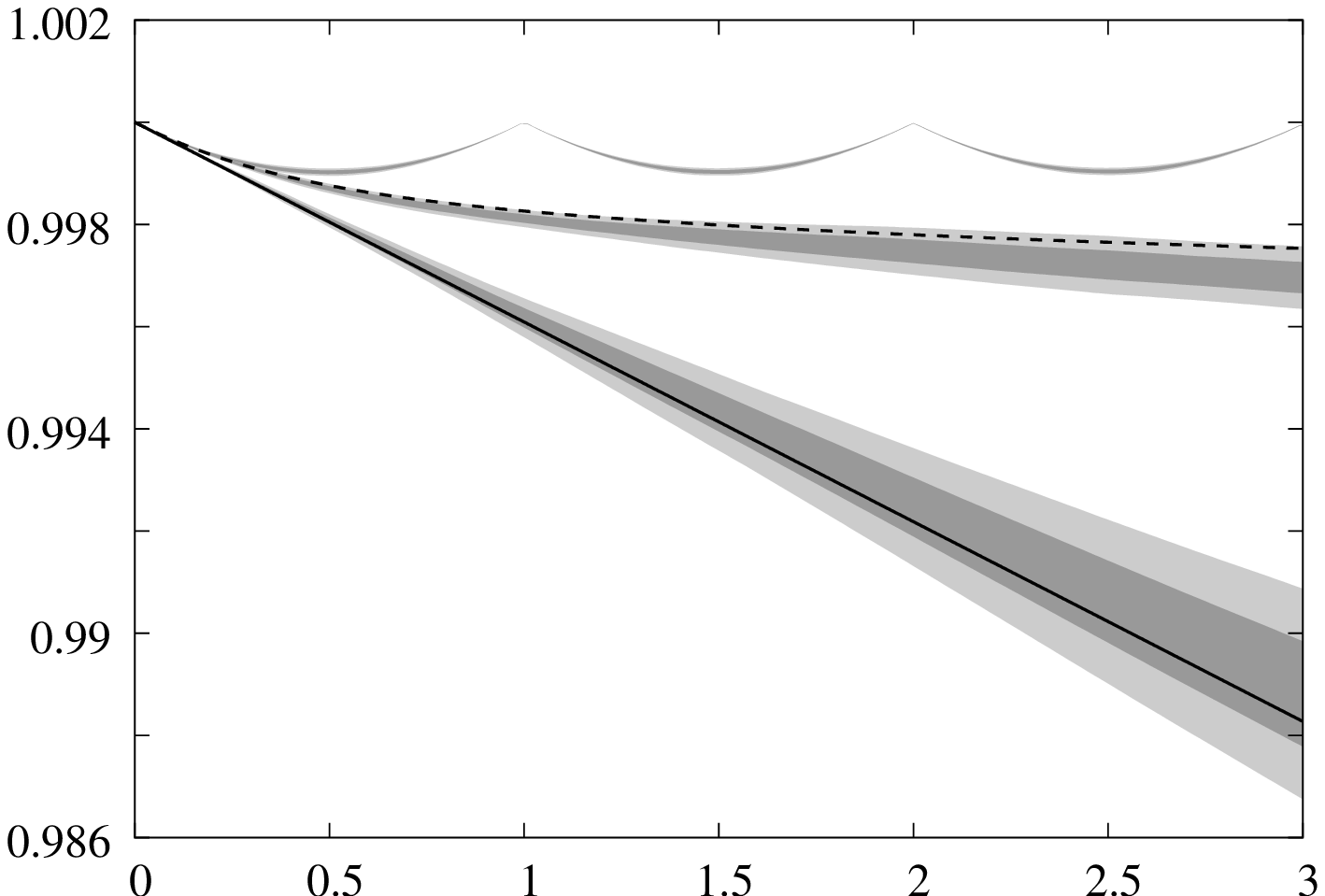}}
\put(230,20){\includegraphics[scale=0.5]{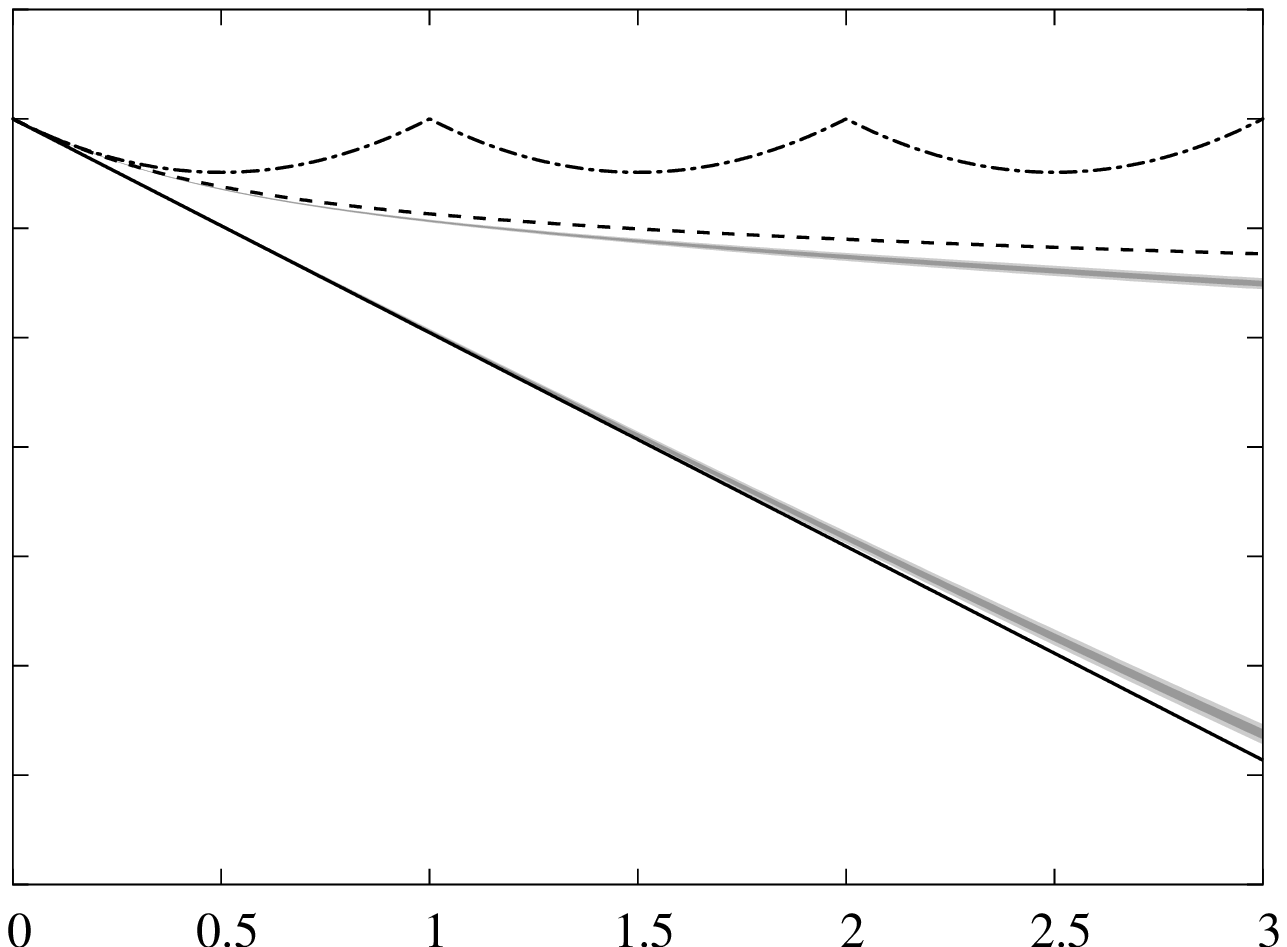}}
\put(190,145){\makebox(0,0){(a)}}
\put(390,145){\makebox(0,0){(b)}}
\put(133,10){\makebox(0,0){$t$}}
\put(324,10){\makebox(0,0){$t$}}
\put(10,90){\makebox(0,0){\begin{sideways} $\la F(t)\ra$\end{sideways}}}
\end{picture}
\caption{\label{f_F1} Fidelity for $N=100, \lambda=0.01$; the initial states 
are eigenstates of $H_0$. (a) The thick solid and dashed lines show the
linear response result~(\ref{F:lresp}) for the Poisson case and the GOE case, 
respectively. The numerical results (shaded areas) are obtained from $10$ 
small samples of size $n_{\rm run}= 1000$, each. The picket-fence case is
also included. The total average of all samples lies in the centre of the 
gray bands, while their borders are given by plus/minus one standard deviation
(dark gray), and plus/minus two standard deviations (light gray). (b) As in 
(a) but for $10$ large samples of size $n_{\rm run}= 50\, 000$. In the 
picket-fence case, the theoretical result is plotted (thick dash-dotted line) 
instead of the numerical data.}
\end{center}
\end{figure}

In figure~\ref{f_F1} we show the fidelity decay for states which are initially
eigenstates of $H_0$. Due to the cancellation of the $t^2$-terms, the 
fidelity decays much slower, and the effects of the spectral correlations are 
more pronounced. For
example, for the picket-fence spectrum, we get practically complete revivals
at integer multiples of the Heisenberg time, while in the Poisson case, the
fidelity decays linearly. Note that exact revivals can occur in the
perturbative limit, only. There the eigenvectors of $H$ are arbitrarily well 
approximated by the eigenvectors of $H_0$. The statistical deviations seem to 
be much smaller in the picket-fence case; see figure~\ref{f_F1}(a). However, 
this is simply because the fidelity is closer to one. We checked that the 
relative error for the correlation integral, remains comparable for all three 
cases. 

In figure~\ref{f_F1}(b)  one can see a small difference between the 
theoretical linear response result and the numerics for both, the Poisson and 
the GOE case. This time, it does not help to exponentiate the linear response 
result. Here, the fidelity decay is too slow to produce an observable effect. 
Note also that the deviations 
differ in sign. While the theory underestimates the fidelity decay in the GOE 
case, it overestimates the decay in the Poisson case. This points at a more
complicated behaviour of the fidelity (as compared to the fidelity amplitude). 
In the picket-fence case, by contrast, the difference between theory and
numerical result is much too small to be resolved. Due to the fact that the 
statistical fluctuations have also diminished, we decided not to show the
numerics at all. 

\begin{figure}
\begin{center}
\setlength{\unitlength}{1pt}
\begin{picture}(445,157)
\put(18,20){\includegraphics[scale=0.5]{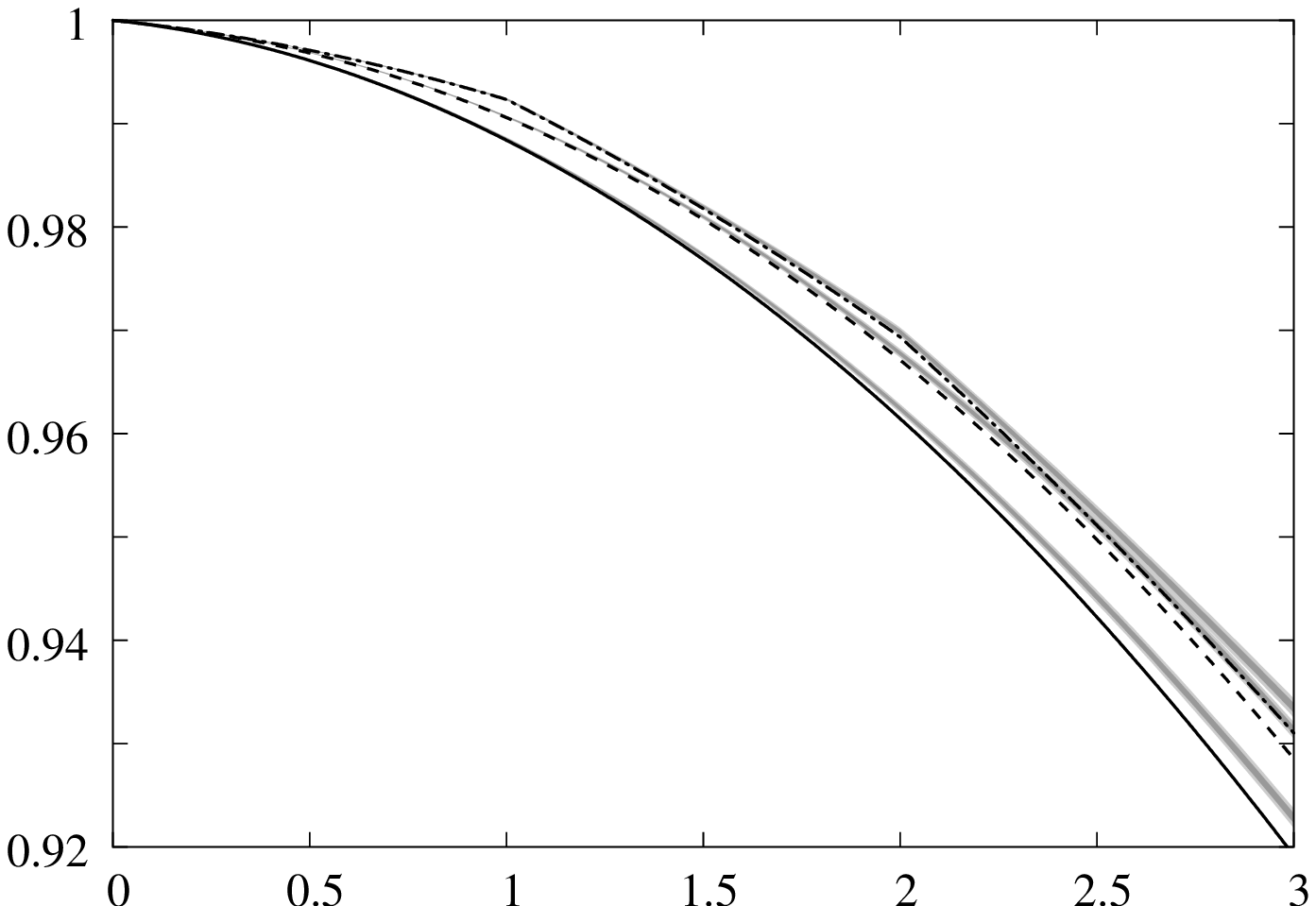}}
\put(236,20){\includegraphics[scale=0.5]{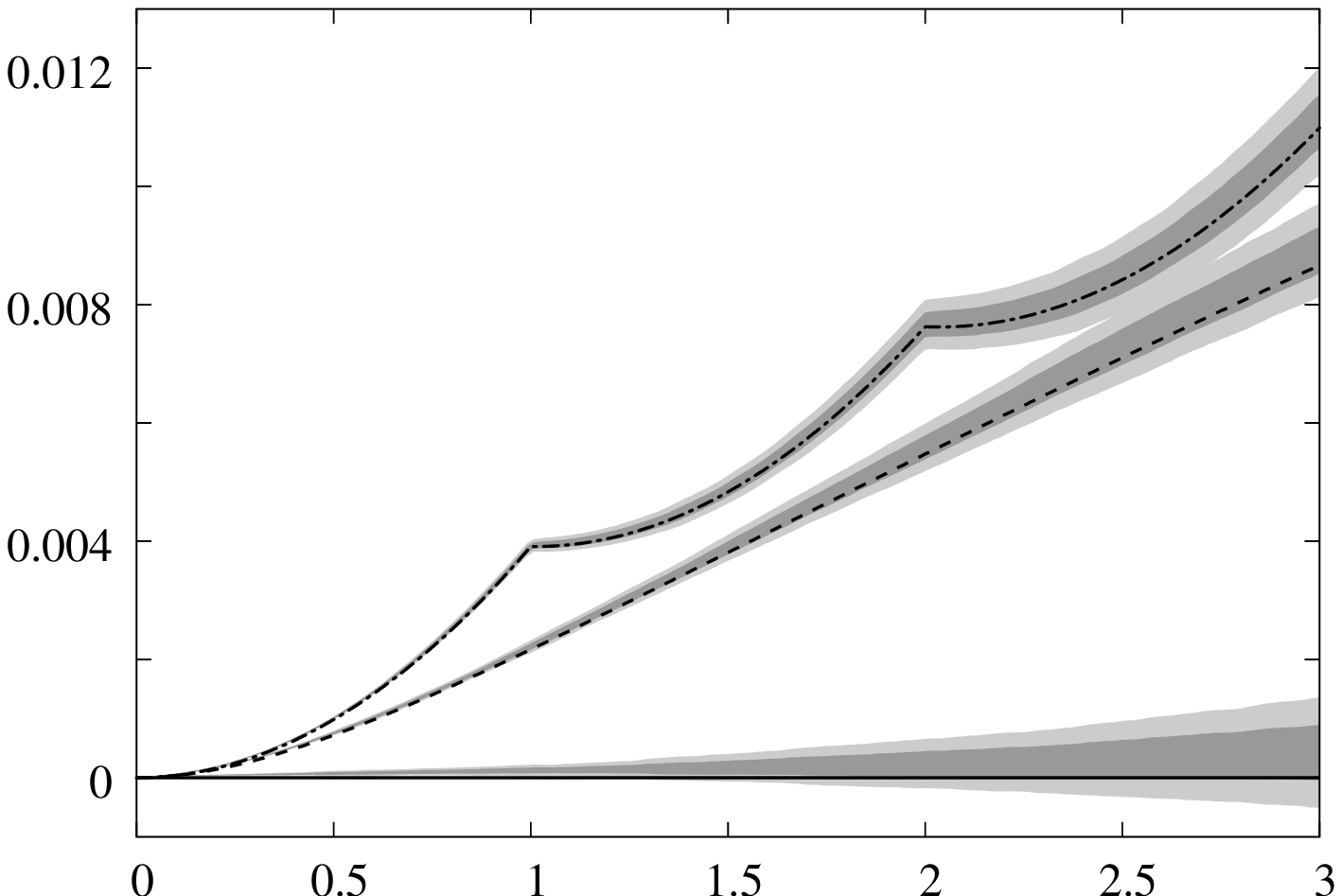}}
\put(200,145){\makebox(0,0){(a)}}
\put(400,145){\makebox(0,0){(b)}}
\put(128,10){\makebox(0,0){$t$}}
\put(353,10){\makebox(0,0){$t$}}
\put(8,90){\makebox(0,0){\begin{sideways} $\la F(t)\ra$\end{sideways}}}
\put(229,90){\makebox(0,0){\begin{sideways}
               $\la F(t)\ra - \la F(t)\ra_{\rm Poi}$\end{sideways}}}
\end{picture}
\caption{\label{f_F2} Fidelity for $N=100, \lambda=0.01$; random initial 
states. Results for $10$ small samples of size $n_{\rm run}= 1000$. The
numerical results are represented by shaded bands, where the significance of
the gray scales is the same as in figure~\ref{f_F1}. (a) Data for the Poisson
case, the GOE-case, and the picket-fence case. The thick lines give the 
theoretical results for the Poisson case (thick solid line), the GOE-case 
(thick dashed line), and the picket-fence case (dash-dotted line). (b) As in 
(a) but with the Poisson theory subtracted (linear response, packed into an 
exponential).}
\end{center}
\end{figure}

In figure~\ref{f_F2} we show the behaviour of the fidelity for random initial
states. As expected, the fidelity behaves very similar to the fidelity 
amplitude in this case ({\it cf.} figures~\ref{f_A1}~and~\ref{f_A2}).
Its decay is again dominated by the $t^2$-term, and the effect of the 
correlations in the spectrum of $H_0$ has decreased, as compared to 
figure~\ref{f_F1}. Again, it makes sense to exponentiate the linear 
response result, in analogy to equation~(\ref{A:eresp}).

\begin{figure}
\begin{center}
\setlength{\unitlength}{1pt}
\begin{picture}(445,157)
\put(18,20){\includegraphics[scale=0.5]{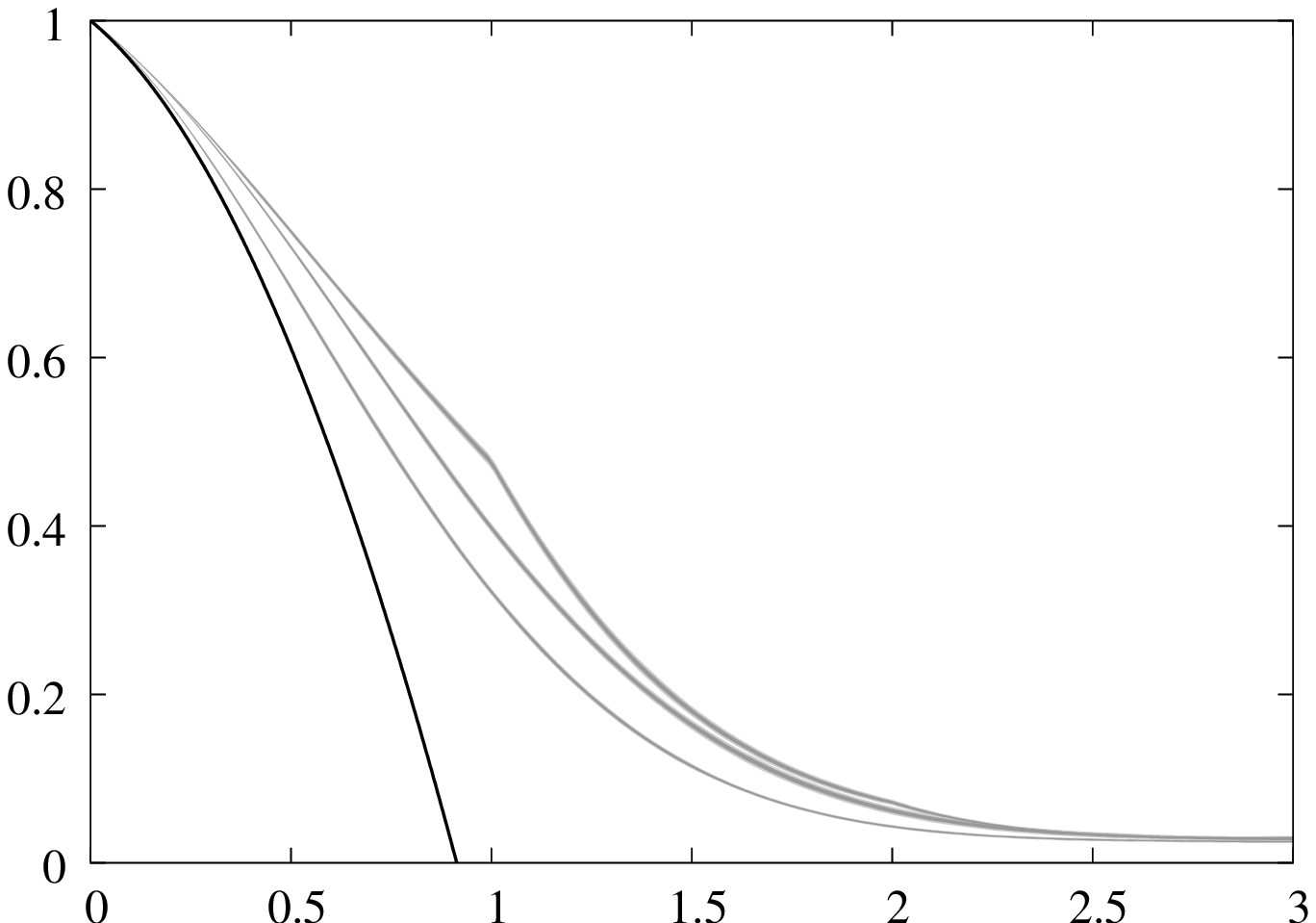}}
\put(242,20){\includegraphics[scale=0.5]{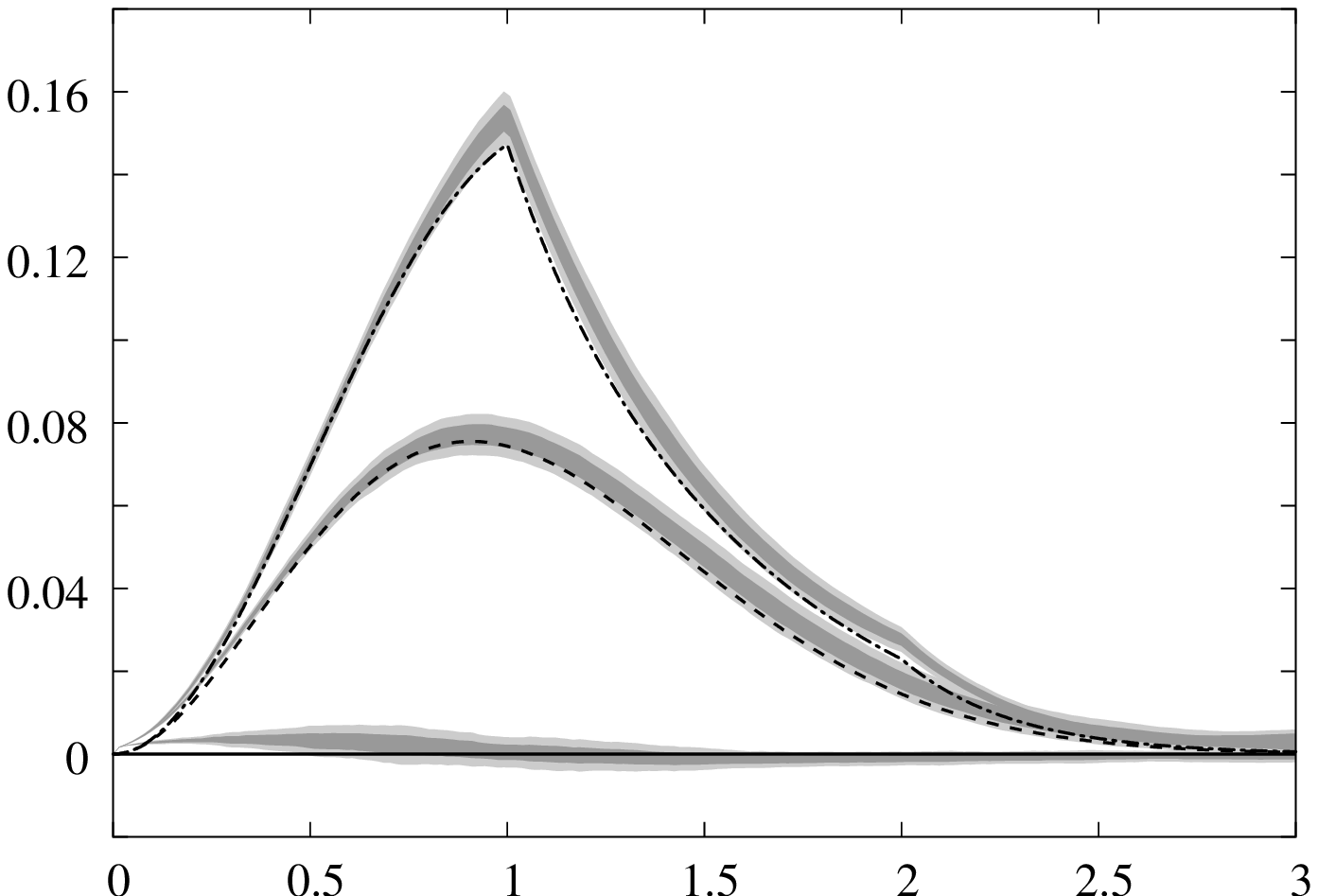}}
\put(200,145){\makebox(0,0){(a)}}
\put(424,145){\makebox(0,0){(b)}}
\put(126,10){\makebox(0,0){$t$}}
\put(354,10){\makebox(0,0){$t$}}
\put(8,90){\makebox(0,0){\begin{sideways} $\la F(t)\ra$\end{sideways}}}
\put(230,90){\makebox(0,0){\begin{sideways} 
               $\la F(t)\ra - \la F(t)\ra_{\rm Poi}$\end{sideways}}}
\end{picture}
\caption{\label{f_F3} Fidelity for $N=100, \lambda=0.1$; random initial states.
Results for $10$ small samples of size $n_{\rm run}= 1000$. The numerical
results are represented by shaded bands, as in figure~\ref{f_F1}. (a) Numerical
results for the Poisson case, the GOE-case, and the picket-fence case. The
thick solid line shows the pure linear response result for the Poisson case. 
(b) as in (a), but with the Poisson theory subtracted (linear response, packed 
into an exponential). The theoretical results, using equation~(\ref{F:hofx}), 
are shown for the Poisson case (thick solid line), GOE-case (thick dashed 
line), and the picket-fence case (thick dash-dotted line).}
\end{center}
\end{figure}

Figure~\ref{f_F3} shows the fidelity decay for random initial states in the
crossover regime, $\lambda = 0.1$, between perturbative and golden rule decay.
According to the discussion of the figures~\ref{corrint-1}~and~\ref{crossover},
we expect a stronger effect of the spectral correlations in this regime.
Note {\it e.g.} at the Heisenberg time, $t = 1$, the relative differences 
between Poisson and GOE as well as between GOE and picket-fence, are of the 
order of $10\%$, each. The theoretical curve in figure~\ref{f_F3}(a) is the
pure linear response result. It is obviously not sufficiently accurate. To
improve, we use exponentiation, but in addition we also take into account a 
rather trivial finite size effect: Namely, as will be shown in the following 
section: $\lim_{t\to\infty} F(t) = c_\infty \approx 3/(N+2)$. Therefore, 
instead of the simple exponentiation, we use
\begin{equation}
h(x)= (1-c_\infty)\; \rme^{-x/(1-c_\infty)} + c_\infty \; , \quad 
c_\infty \approx \frac{3}{N+2} \; .
\label{F:hofx}\end{equation}
This phenomenological formula describes the numerical results quite well, as
can be seen in figure~\ref{f_F3}(b).

\nosections

\noindent
In all the cases considered, we find that the spectral correlations (or more 
precisely, the level repulsion) tend to attenuate fidelity decay. 
Qualitatively, 
this can be understood from standard perturbation theory~\cite{Messiah79}, 
which relates changes in the eigenvector components of $H$ to the matrix
elements of the perturbation $\lambda V_{\alpha\beta}$, weighted with the 
inverse level distance $(\oE_\alpha - \oE_\beta)^{-1}$. Accordingly, the
fidelity decay is slowest in the picket-fence case. In the GOE case (which
may be associated with chaotic dynamics) it is somewhat faster, while in the 
Poisson case (characteristic of integrable dynamics) it is fastest. In this 
sense, we can confirm that fidelity is more likely to be maintained, if 
the system's dynamics is chaotic rather than integrable. This, at first sight
counter-intuitive result, has been obtained previously by a number of 
authors~\cite{ProZni02,ProZni01,ProSel02},
although along quite different lines of argument.

\subsection{Fidelity distributions}

In particular for eigenstates of $H_0$, we saw a dramatic effect of the 
spectral correlations on the decay of the average fidelity. As it turned out,
sufficiently strong correlations can practically stop the decay of 
$\la F(t)\ra$ at values quite close to one (cf. figure~\ref{f_F1}). However,
the large sample-to-sample fluctuations in this case make us suspect that
for individual systems this stabilisation effect might be reduced, or not 
even observable. However, in the case of the fidelity amplitude, we could
actually prove that if the IPR of the initial state is only small enough,
then effects of the spectral correlations are observable even for individual
systems. We expect that the situation will be similar in the case of the 
fidelity itself. There will be a trade off between quantum ergodicity and the
size of spectral correlations. Besides of its own interest, we study the
fidelity distributions $P(F(t))$ to clarify this point.
We generate the required histograms at
certain equidistant time instances. For each histogram, the interval $(0,1]$ 
is divided into $50$ boxes. Then, for a certain sample of $n_{\rm run}$ systems
we count the frequencies with which the fidelity for an individual system, at 
a given time, falls into one of those boxes. After proper normalisation the 
histograms are lined up along the time axis, which results in three 
dimensional plots.

\begin{figure}
\begin{center}
\setlength{\unitlength}{1pt}
\begin{picture}(445,215)
\put(0,70){\includegraphics[scale=0.5]{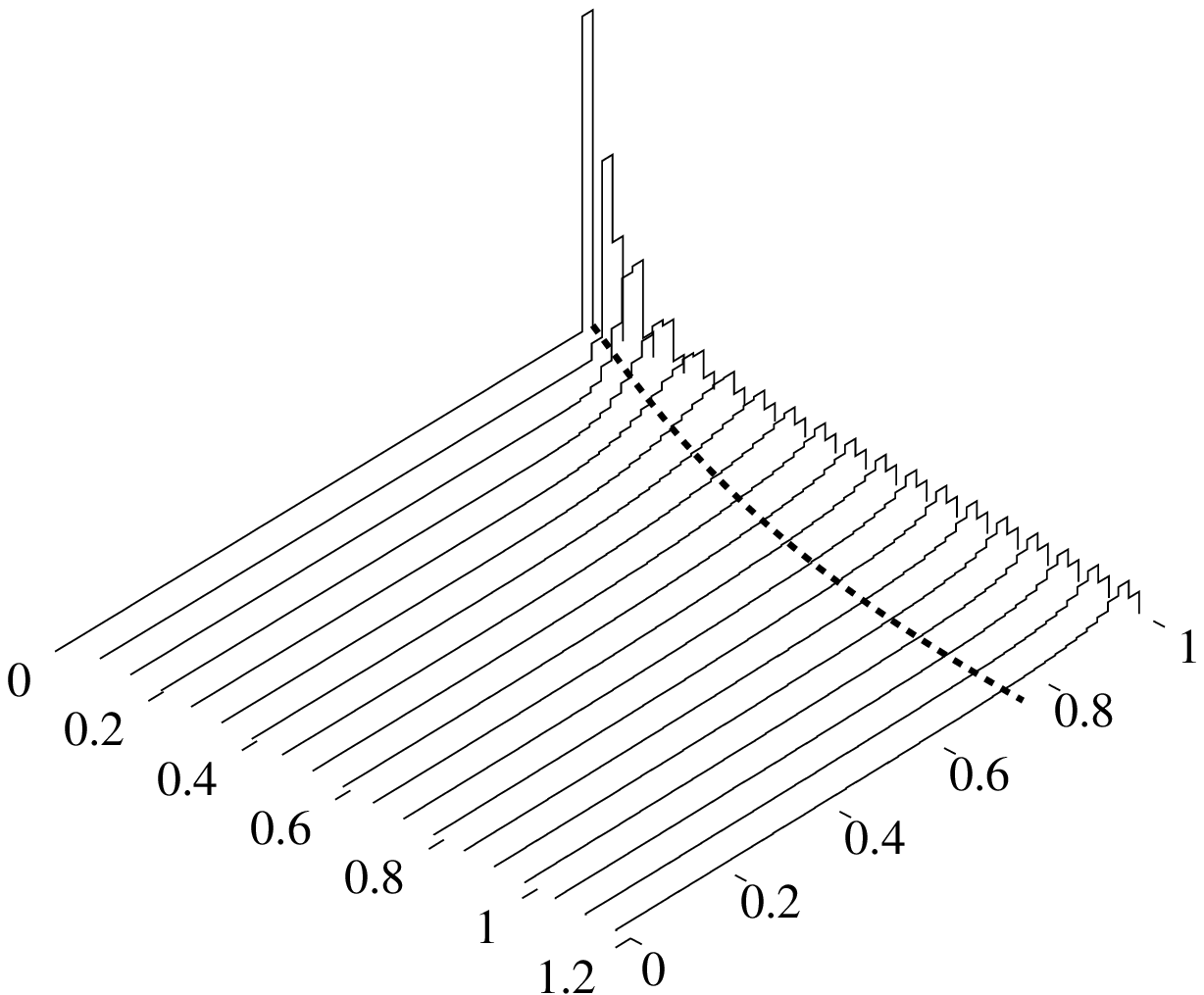}}
\put(120,10){\includegraphics[scale=0.5]{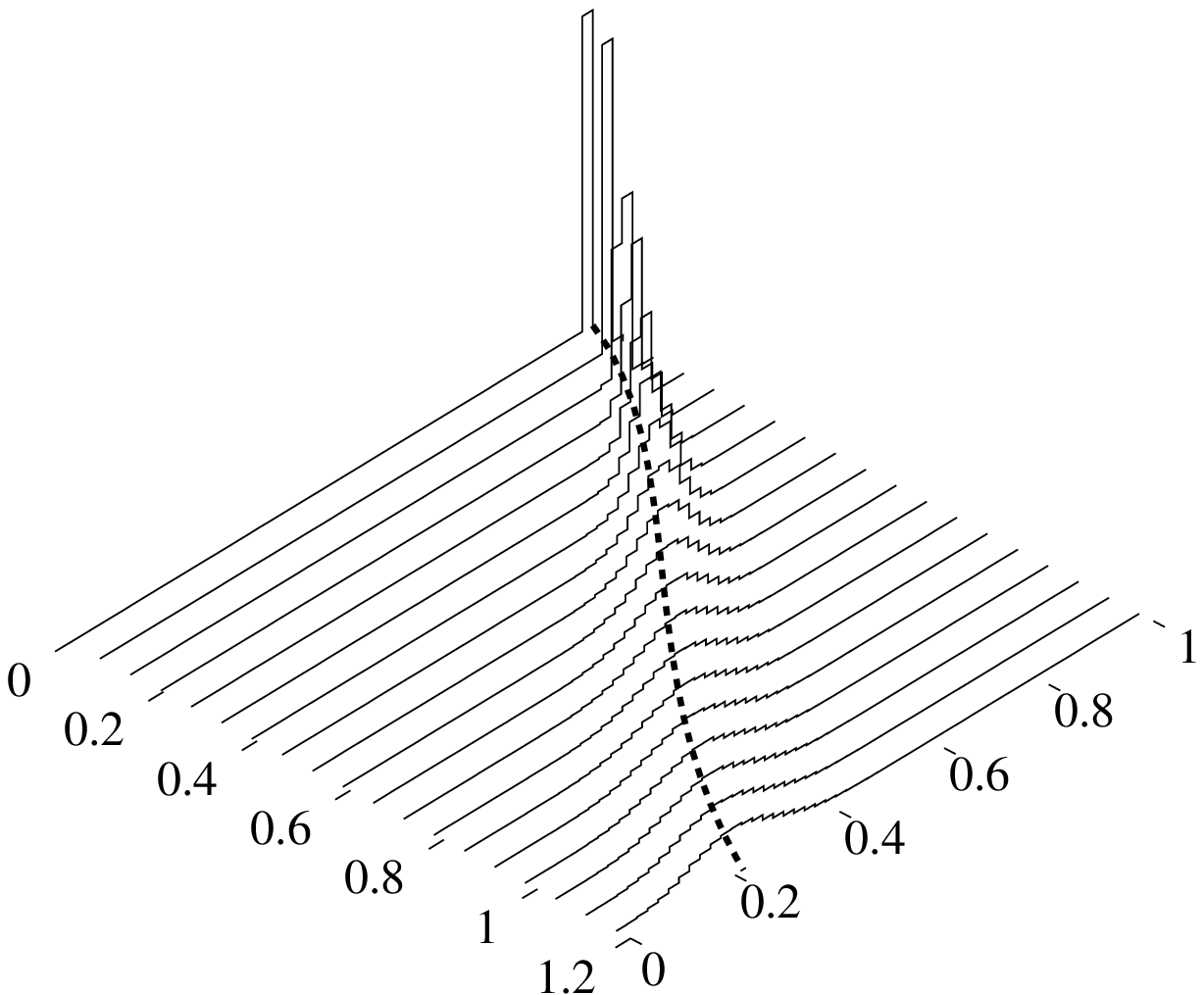}}
\put(260,70){\includegraphics[scale=0.5]{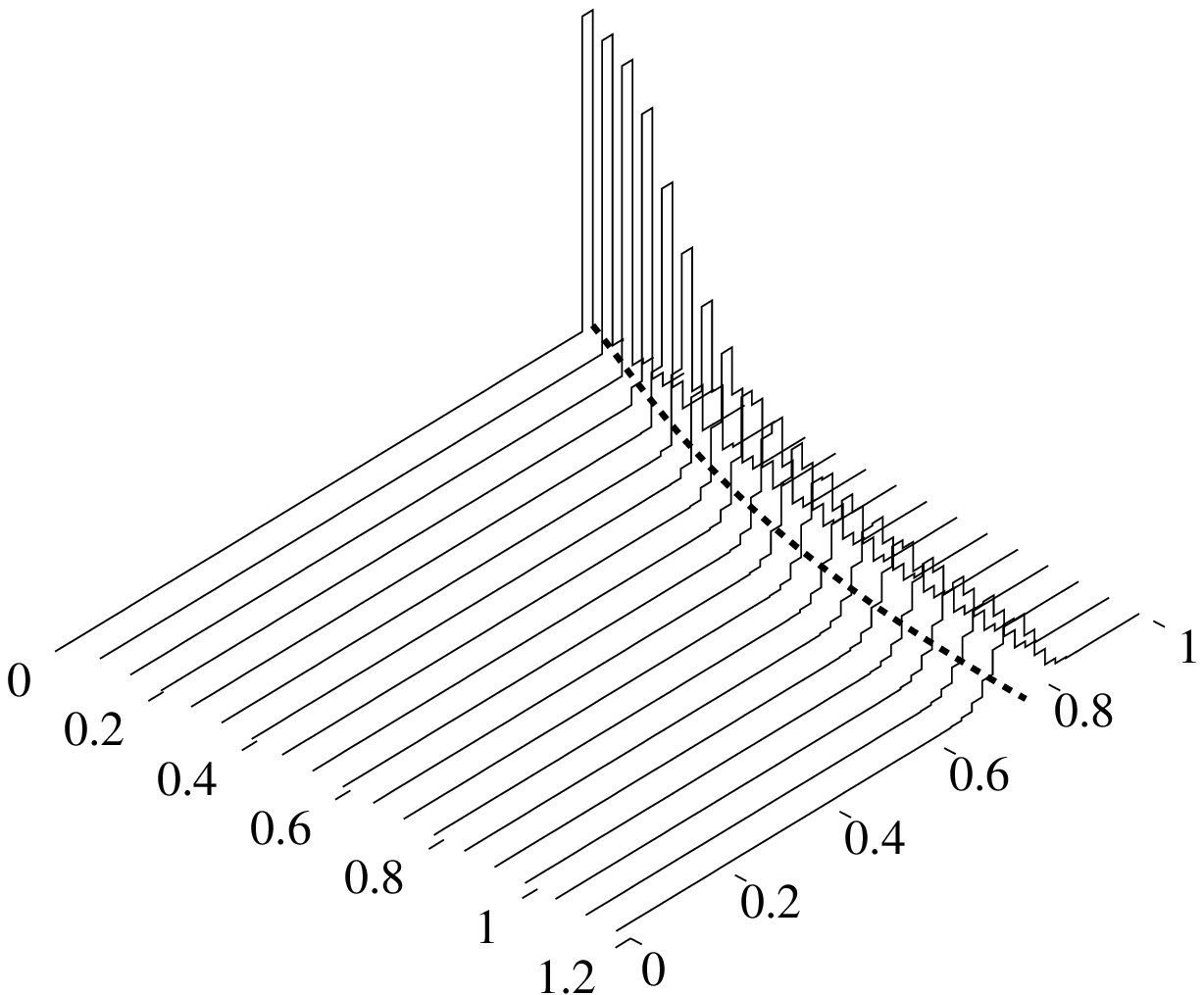}}
\put(120,180){\makebox(0,0){(a)}}
\put(230,135){\makebox(0,0){(b)}}
\put(380,180){\makebox(0,0){(c)}}
\put(40,85){\makebox(0,0){$t$}}
\put(160,25){\makebox(0,0){$t$}}
\put(280,25){\makebox(0,0){$F(t)$}}
\put(420,85){\makebox(0,0){$F(t)$}}
\end{picture}
\caption{\label{f_FN2-1} Fidelity distribution for $N=100, \lambda=0.1$ for
the Poisson case. Histogram as a function of time (solid lines), for initial 
eigenstates (a), for random initial states (b), and for the average over all 
possible initial eigenstates (c). The distributions are normalised, such that 
{\it e.g.} at $t=0$ the height of the bar is $50$, because the widths of the 
histogram boxes is $1/50$. The average fidelity is plotted as a thick dashed
line in the $xy$-plane.}
\end{center}
\end{figure}

In figure~\ref{f_FN2-1} we focus on the qualitative features of the
fidelity distribution for different choices of the initial states. For 
simplicity we show the Poisson case, though note that our observations are 
equally valid for the GOE case. In all three panels we show the fidelity
distribution as it evolves in time, together with the behaviour of the average
fidelity (thick dashed line in the $xy$-plane). For $H_0$-eigenstates, 
panel~(a), the maximum of the distribution remains very close to one, while
its average continues to decay (very slowly, though). In this case, linear 
response predicts a purely linear decay, which can be observed at small 
times. However, for larger times $t \gtrsim 0.3$ the average fidelity starts
to saturate, an effect we cannot describe theoretically (as we noticed already
in figure~\ref{f_F1}, exponentiation does not help, here). The separation
between the average and the maximum of the fidelity distribution is due to 
the development of a long range tail.

For random initial states, panel~(b), the fidelity distribution is rather
symmetric around its average value. Its shape looks almost Gaussian. We 
checked that the width of the distribution decreases roughly proportional to 
$1/\sqrt{N}$. This means that we have true quantum ergodicity for the
fidelity of a random initial state. For sufficiently high dimension of the
Hamiltonian, the sample-to-sample fluctuations can be made arbitrarily small.

In panel~(c), we show the fidelity distribution, where, for each individual 
system, we have averaged over all available $H_0$-eigenstates as initial 
states. This also reduces the sample-to-sample fluctuations drastically. 
Though,
qualitatively, the width of the fidelity distribution behaves similar to the 
case with random initial states, it is noticeably smaller. Of course, the 
average fidelity shows here exactly the same behaviour as in panel~(a).

\nosections

\begin{figure}
\begin{center}
\setlength{\unitlength}{1pt}
\begin{picture}(445,235)
\put(20,20){\includegraphics[scale=0.36]{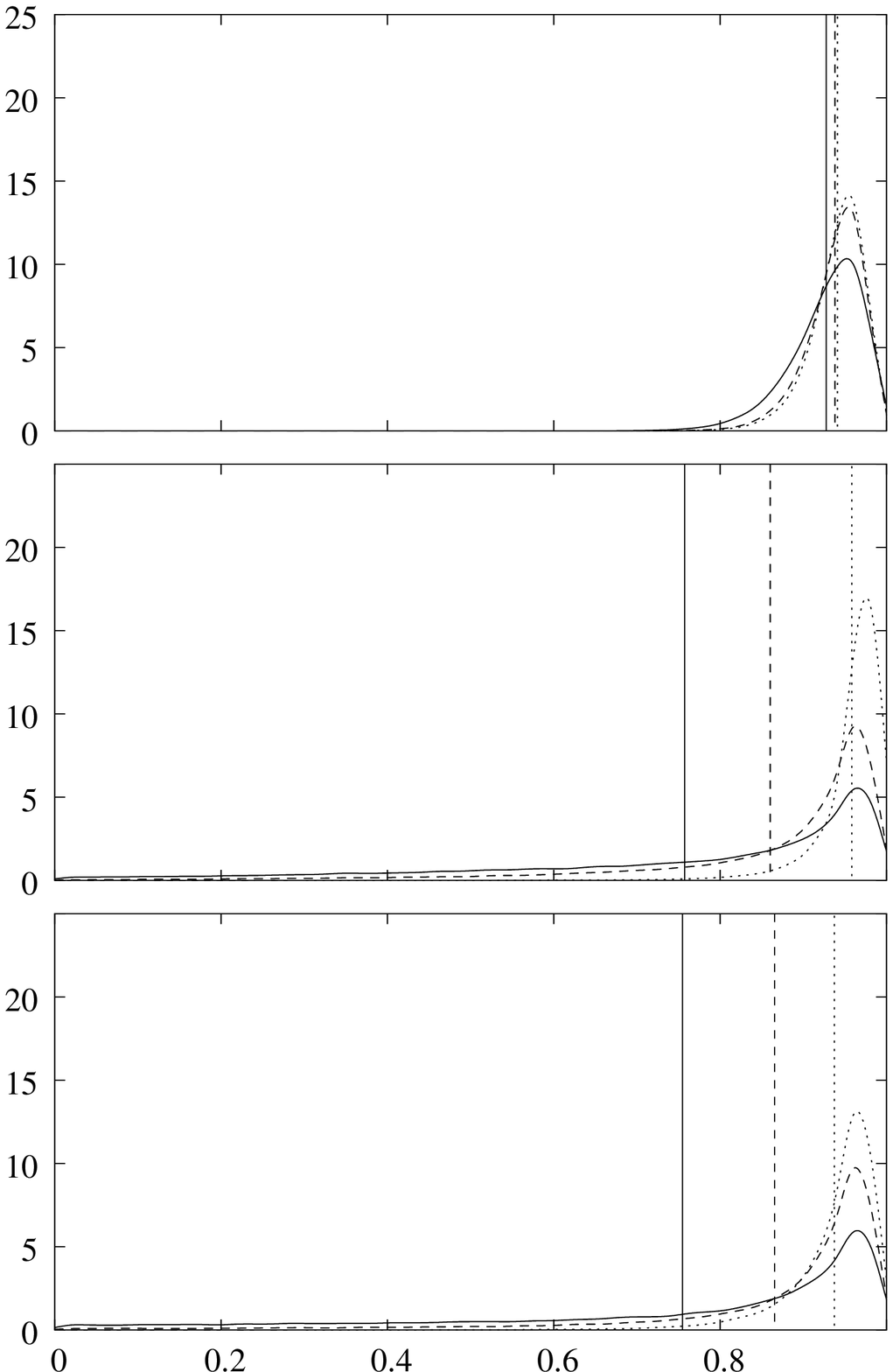}}
\put(169,20){\includegraphics[scale=0.36]{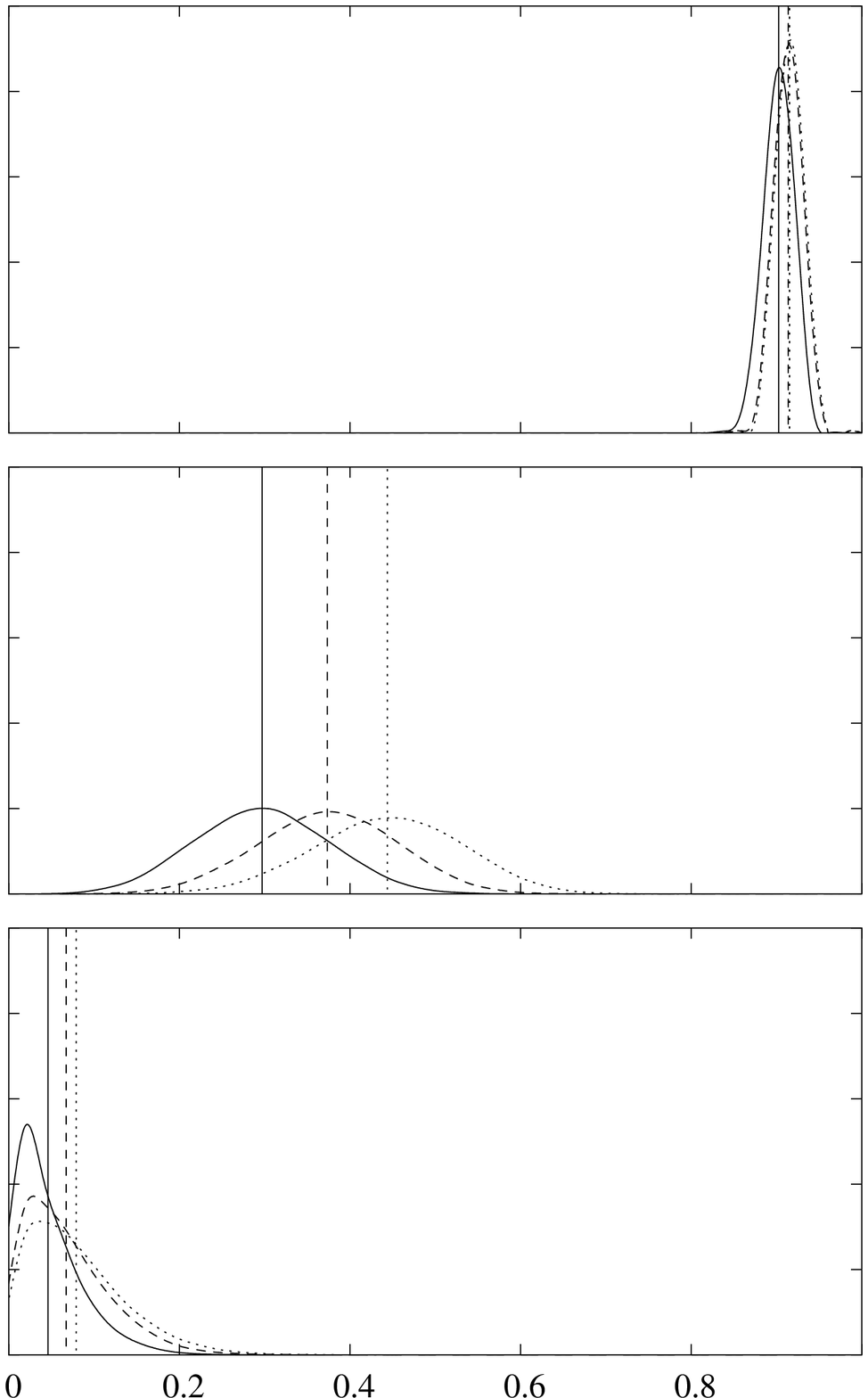}} 
\put(310,20){\includegraphics[scale=0.36]{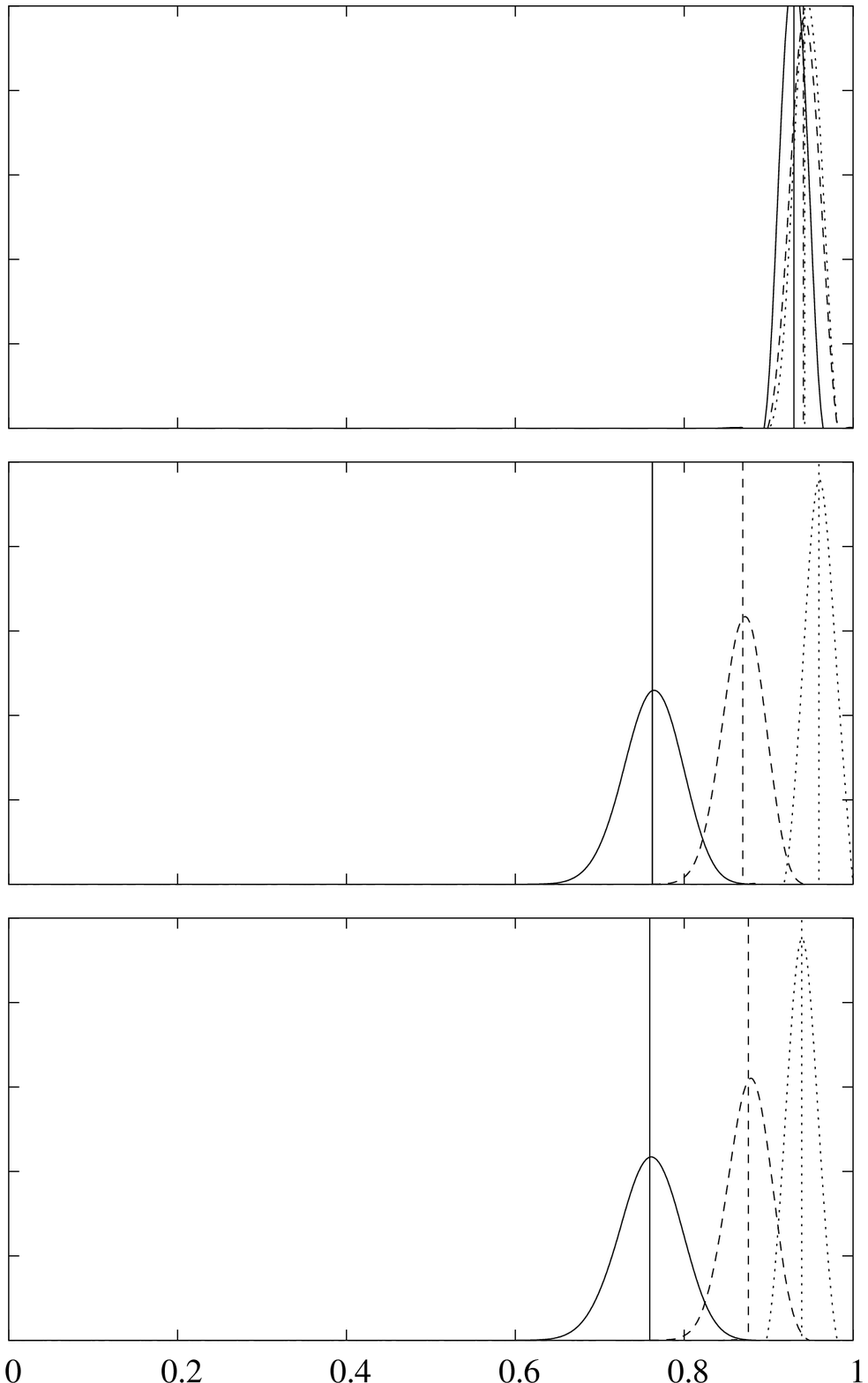}}
\put(46,220){\makebox(0,0){(a)}}
\put(192,220){\makebox(0,0){(b)}}    
\put(325,220){\makebox(0,0){(c)}}
\put(46,150){\makebox(0,0){(d)}}
\put(192,150){\makebox(0,0){(e)}}    
\put(325,150){\makebox(0,0){(f)}}
\put(46,80){\makebox(0,0){(g)}}
\put(192,80){\makebox(0,0){(h)}}     
\put(325,80){\makebox(0,0){(i)}}
\put(235,10){\makebox(0,0){$F(t)$}}  
\put(8,120){\makebox(0,0){\begin{sideways} $P(F(t))$\end{sideways}}}
\end{picture}
\caption{\label{f_FN2-2} Fidelity distributions for $N= 100, \lambda=0.1$ for
the Poisson case (solid line), the GOE case (dashed line), and the 
picket-fence case (dotted line) at three different times: $t=0.195$ 
[(a),(b),(c)], 
$t=0.974$ [(d),(e),(f)], and $t=1.884$ [(g),(h),(i)]. In the left column
[(a),(d),(g)], the initial states were eigenstates of $H_0$, in the middle 
column [(b), (e), (h)] we used random initial states, and in the right column
[(c),(f),(i)] we averaged over all initial eigenstates of $H_0$. The vertical
lines give the average value of the fidelity for the whole sample for the 
Poisson case (solid line), for the GOE case (dotted line), and for the 
picket-fence case (dashed line).}
\end{center}
\end{figure}

\noindent
In figure~\ref{f_FN2-2} we show examples of fidelity distributions for the
Poisson, the GOE, and the picket-fence case, at three different times $t_1$, 
$t_2$, and $t_3$. At a time $t_1= 0.195$ much before the Heisenberg time,
at Heisenberg time $t_2= 0.974$ and at a time much after the Heisenberg time
$t_3= 1.884$. The different panels are arranged into columns, where in the 
left column we used $H_0$-eigenstates as initial states, in the middle column
we used random initial states, and in the right column, we averaged over all
$H_0$-eigenstates. 

In the left column we clearly see the development of a long range tail towards
low fidelities, which causes the separation between the average and the maximum
of the fidelity distribution. Though the tail is more pronounced in the
Poisson case, it is still important in the GOE case, but may be somewhat less
important in the picket-fence case. In the middle and the right column, the
fidelity distributions have almost perfect symmetric shapes, at least as long
as the distributions do not come too close to the boundaries zero and one.
In the case where we averaged over the $H_0$-eigenstates (right column), the
widths of the distributions are small enough to allow to distinguish even
individual systems with respect to the different spectral correlations. For
random initial states (middle column), the widths of the fidelity distributions
are not small enough, and one would have to use bigger Hamiltonian matrices
($N\gtrsim 1000$).

\section{\label{T} The fidelity in the long time limit}

Here we will calculate the survival probability, the fidelity amplitude, and
the fidelity in the limit $t\to\infty$ (with the strength of the perturbation
held fixed). One may think that in general, only the eigenvectors are 
important. As will be shown below, this is indeed the case, as long as $H_0$ 
is not constructed from a picket-fence spectrum. 
We use time independent perturbation theory, to obtain 
approximations to the eigenvectors. The details can be found in~\ref{A_Pert}.

Remember that we are using units for time and energy such that the 
Heisenberg time and the mean level spacing are both equal to one 
(see section~\ref{A}). In these 
units, the time dependent Schr\" odinger equation reads:
\begin{equation}
\rmi\partial_t\, \Psi(t) = 2\pi\; (H_0 + \lambda\, V)\; \Psi(t) = 0 \; .
\end{equation}
Suppose we can diagonalise $H= H_0 + \lambda\, V$ exactly. Then, an initial 
state $\ox$ will evolve in time according to
\begin{equation}
\fl u(t) = O\; \Delta(t)\; O^T \; \ox \; , \quad 
\Delta(t) = {\rm diag}\left(\rme^{-2\pi\rmi\, E_\alpha\, t}\right) \; , \quad
O^T\; H\; O = {\rm diag}(E_\alpha) \; .
\end{equation}
We assume that $H_0$ is diagonal, so that we obtain for the evolution in the
interaction picture:
\begin{equation}
\fl x(t) = \Delta_0(t)^{-1}\; u(t) = \Delta_0(-t)\; O\; \Delta(t)\; O^T \; \ox 
\; , \quad
\Delta_0(t) = {\rm diag}(\rme^{-2\pi\rmi\, \oE_\alpha\, t}) \; ,
\end{equation}
where $\oE_\alpha$ are the eigenvalues of $H_0$. Here, 
$\Delta_0(-t)\; O\; \Delta(t)\; O^T$ is the exact echo operator introduced in
equation~(\ref{defH}). We expect, that in the long time limit, all 
dependencies on the eigenvalues of $H$ and $H_0$ will drop out, so that the 
problem consists in finding an appropriate approximation for the eigenvectors, 
{\it i.e.} for the orthogonal matrix $O$.

\nosections

\noindent
The fidelity amplitude can be expressed as the expectation value of the
echo operator with respect to the state $\ox$.
\begin{equation}
\fl f(t) = \la\Delta_0(-t)\; O\; \Delta(t)\; O^T\ra_{\ox}
= \sum_{\alpha\beta\gamma} \ox_\alpha\; \rme^{2\pi\rmi\, \oE_\alpha\, t} \;
    O_{\alpha\beta}\; \rme^{-2\pi\rmi\, E_\beta\, t} \; O_{\gamma\beta}\; 
    \ox_\gamma \; .
\end{equation}
In this expression,the phases in the exponential can never cancel. Therefore, 
the fidelity amplitude, averaged over the perturbation $V$ and over $H_0$, 
vanishes in the limit $t\to\infty$.
\begin{equation}
\lim_{t\to 0}\; \la f(t)\ra_{0,V} = 0 \; .
\end{equation} 
For the fidelity, we obtain a more complicated expression:
\begin{eqnarray}
\fl F(t) = \left| \sum_{\alpha\beta\gamma} \ox_\alpha\; 
    \rme^{2\pi\rmi\, \oE_\alpha\, t} \; O_{\alpha\beta}\; 
    \rme^{-2\pi\rmi\, E_\beta\, t} \; O_{\gamma\beta}\; \ox_\gamma \right|^2
\nonumber\\
\lo\to \sum_{\alpha\alpha'}\sum_\beta \sum_{\gamma\gamma'} \ox_\alpha\; 
   \ox_{\alpha'}\; \rme^{2\pi\rmi (\oE_\alpha - \oE_{\alpha'})\, t}
   \; O_{\alpha\beta}\; O_{\alpha'\beta}\; O_{\gamma\beta}\; 
   O_{\gamma'\beta}\; \ox_\gamma\; \ox_{\gamma'} \; .
\end{eqnarray}
Here, the arrow indicates that we discarded already those terms containing
phases of the form $\exp[-2\pi\rmi(E_\beta - E_{\beta'})\, t]$, as they cannot
contribute to the long time limit of the average fidelity.

If $\ox$ is an eigenstate of $H_0$, $\ox_\beta = \delta_{\alpha\beta}$, then 
the fidelity as a function of time coincides with the survival probability, 
as we have seen in the beginning of section~\ref{F}. For the limit value of 
both quantities, we thus get:
\begin{equation}
F^e_\infty = \lim_{t\to\infty} \la F(t)\ra_{0,V} 
           = \lim_{t\to\infty} \la S(t)\ra_{0,V} 
           = \sum_\xi \la O_{\alpha\xi}^4\ra_{0,V} 
           = \la{\rm ipr}(O^T\, \ox)\ra_{0,V} \; .
\label{T:Feinfty}\end{equation}
This is the inverse participation ratio of the local density of states
({\it i.e.} the projections of the $H_0$-basis state onto the eigenbasis of 
$H$). By contrast, if $\ox$ is a random state, 
$\la F(t)\ra_{0,V} \to F^r_\infty$, where:
\begin{eqnarray}
\fl F^r_\infty = \sum_{\alpha\ne\gamma} \sum_\beta \lla\ox_\alpha^2\; 
   O_{\alpha\beta}^2\; O_{\gamma\beta}^2\; \ox_\gamma^2\rra_{0,V}
  + \sum_{\alpha\ne\alpha'}\sum_\beta \lla\ox_\alpha^2\; \ox_{\alpha'}^2\;
    \rme^{2\pi\rmi (\oE_\alpha - \oE_{\alpha'})\, t}\; O_{\alpha\beta}^2\;
    O_{\alpha'\beta}^2\rra_{0,V} \nonumber\\
  + \sum_{\alpha\ne\gamma} \sum_\beta \lla\ox_\alpha^2\; \ox_\gamma^2\;
    \rme^{2\pi\rmi (\oE_\alpha - \oE_\gamma)\, t}\; O_{\alpha\beta}^2\; 
    O_{\gamma\beta}^2\rra_{0,V} \nonumber\\
\fl\qquad= \frac{1}{N(N+2)}\sum_\beta \lla \sum_{\alpha\ne\gamma} 
    O_{\alpha\beta}^2\; O_{\gamma\beta}^2 \left( 1 + 2\; 
    \rme^{2\pi\rmi (\oE_\alpha - \oE_{\alpha'})\, t}\right)  + 
    3\sum_\alpha O_{\alpha\beta}^4 \rra_{0,V} \; .
\end{eqnarray}
For long times, the time dependence survives only in the picket-fence case.
Then, at integer values of $t$, the fidelity reaches exactly 
the value of the survival probability. 
In between, the exponentials give zero on the average. In all 
other cases, where the spectrum of $H_0$ is sufficiently random, the 
exponentials give always zero. To proceed, we will now calculate the limit 
value of the fidelity (disregarding the exponential). In this case: 
\begin{eqnarray}
F^r_\infty &= \frac{1}{N(N+2)}\sum_\xi \lla 
     \sum_{\alpha\ne\beta} O_{\alpha\xi}^2\; O_{\beta\xi}^2 +
     3\sum_\alpha O_{\alpha\xi}^4 \rra_{0,V} \nonumber\\ 
&= \frac{1}{N+2}\left( 1 + \frac{2}{N}\sum_{\alpha\xi} 
  \la O_{\alpha\xi}^4\ra_{0,V} \right) \; ,
\end{eqnarray}
due to $\sum_{\alpha\beta} O_{\alpha\xi}^2\, O_{\beta\xi}^2 = 1$. Hence,
both results are linearly related:
\begin{equation}
F^r_\infty = \frac{1 + 2\; F^e_\infty}{N+2} \; .
\end{equation}

\subsection*{Perturbation theory for $F^e_\infty$}

Here we consider the long time limit $F^e_\infty$ of the fidelity for initial
$H_0$-eigenstates, as defined in equation~(\ref{T:Feinfty}). For the GOE case,
as well as for the Poisson case, the standard (Rayleigh-Schr\" odinger) 
perturbation theory~\cite{Messiah79} leads to divergences. To obtain 
meaningful results one had to use degenerate perturbation theory, as 
in~\cite{CerTom03}. Unfortunately, this is quite involved in our case, where 
powers of matrix elements occur up to order four.
Therefore we restrict our analytical treatment to the picket-fence case. We
assume that the initial state is taken from the centre of the spectrum. As 
the spectrum is deterministic, only the average over the GOE perturbation 
matrix~$V$ has to be performed. All averages required are of the form 
$\la O^2_{\xi\beta}\, O^2_{\mu\beta}\ra_V$. They are calculated 
in~\ref{A_Pert}, using the standard perturbation theory. Up to forth order 
we obtain:
\begin{equation}
\la O^4_{\alpha\alpha}\ra_V = 1 - 2\lambda^2\; \frac{\pi^2}{3}
     + \lambda^4\; \frac{\pi^4}{5} \qquad
\la O^4_{\xi\alpha}\ra_V = \frac{3\; \lambda^4}{(\alpha-\xi)^4} \; ,
\end{equation}
(it is assumed that $\alpha\ne\beta$) and therefore
\begin{equation}
F_\infty^e = 1 - 2\lambda^2\; \frac{\pi^2}{3} + 
     4\lambda^4\; \frac{\pi^4}{15} + \Or(\lambda^6)\; .
\label{T_Finf:PIF}\end{equation}

\begin{figure}
\begin{center}
\setlength{\unitlength}{1pt}
\begin{picture}(261,185)
\put(20,20){\includegraphics[scale=0.6]{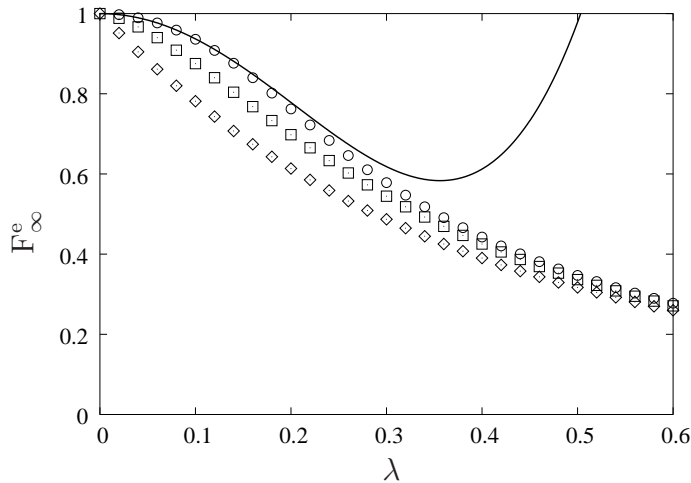}}
\put(148,10){\makebox(0,0){$\lambda$}}
\put(10,100){\makebox(0,0){\begin{sideways} $\rm F^e_\infty$\end{sideways}}}
\end{picture}
\caption{\label{f_ipn} The average fidelity in the large time limit for 
initial eigenstates of $H_0$ ($N=100$). Numerical results for large samples 
$n_{\rm run}= 50\, 000$, for the Poisson case (\opendiamond), the GOE 
(\opensquare), and the picket-fence case (\opencircle). Perturbation
theory, equation~(\ref{T_Finf:PIF}), for the picket-fence case (solid line).}
\end{center}
\end{figure}

Figure~\ref{f_ipn} shows $F^e_\infty$ as a function of the perturbation 
strength $\lambda$ for the three different spectral ensembles under 
consideration. In general the (asymptotic) behaviour of $F^e_\infty(\lambda)$
for small $\lambda$ strongly depends on the degree of level repulsion in the
respective spectrum. Our numerical results range from an apparently linear 
decrease in the Poisson case up to a quadratic decrease in the picket-fence 
case, while for the GOE case we obtain some intermediate behaviour. For large
$\lambda$ all curves for $F^e_\infty(\lambda)$ seem to merge into a single one.
In the picket-fence case we observe that the perturbative result, 
equation~(\ref{T_Finf:PIF}), describes the behaviour at small $\lambda$ very 
well.

\section{\label{C}Conclusions}

In this paper, we have developed and analysed a random matrix
model for echo dynamics in a chaotic system. The fidelity has been computed
in the linear response approximation, and a transition from a linear decay 
to a quadratic one at times of the order of half the Heisenberg time      
has been found. It is easily verified that in the limit of small perturbations
one obtains quite generally a Gaussian decay. By exponentiating our
linear response result, we have obtained a theoretical description which 
reproduces the Gaussian decay in this limit. By construction, this procedure 
provides a fair approximation for the perturbative and the golden rule regime. 
However, by comparing with recent fidelity studies for a deterministic chaotic 
system~\cite{CerTom03}, we could show that our approximation is
accurate in the crossover regime also.

The main new ingredient in our approach is the dependence of the fidelity
decay on the spectral two-point form factor of the underlying unperturbed
Hamiltonian. This lead us to systematically study the effects of spectral
correlations on the fidelity decay. Besides the GOE type fluctuations which
we expect in classically chaotic systems, we also investigated random 
spectra (no correlations) and picket-fence spectra (maximal correlations) 
as the two extreme cases.

In general, we have found that spectral correlations ({\it i.e.} level
repulsion) tend to inhibit fidelity decay. However, the effect is usually
quite small. It becomes big only if we choose eigenstates of the 
unperturbed Hamiltonian as initial states. Though the correlation effects
are really dominant in this case, quantum ergodicity is then lost. This 
means that the sample-to-sample fluctuations are as large as the 
fidelity signal itself, so that it is absolutely necessary to perform an
ensemble average over many systems, or at least many initial states. In 
order to restore quantum ergodicity, we need to consider initial states with
large IPR's (if expanded in the unperturbed basis). Unfortunately, this 
inevitably leads to much smaller correlation effects. However, these effects
are still observable, even for a single fidelity experiment, provided that
the IPR is sufficiently small ($\sim 10^{-3}$).

Finally, we considered the fidelity in the limit of large times. We have shown
that the result for a random initial state is related in a simple way to the
result for an eigenstate of the unperturbed Hamiltonian. Again we investigated
the effect of spectral correlations on this limit value for the fidelity. 
Using perturbation theory, we could obtain an analytical result for the 
picket-fence case, valid for small perturbations.

\ack
We acknowledge useful discussions with W.~Strunz and F.~Leyvraz. We are
grateful to N.~Cerruti and S.~Tomsovic for providing us with the original 
data from their standard map calculations (figure 7). This work
was supported by the DGAPA (UNAM) grant IN-109000, the CONACyT grant 25192-E, 
as well as the EU Human Potential Program contract HPRN-CT-2000-00156. 
TP acknowledges financial support by Ministry of Education, Science and 
Sports of the Republic of Slovenia, and in part by the US ARO grant 
DAAD19-02-1-0086.

\begin{appendix}

\section{\label{A_Cint} The correlation integral for different statistical
   spectra}

In order to calculate the correlation integral, equation~(\ref{cint}), we need 
the second integral of the two-point form factor. If we denote the first 
integral by $B(t)$, we may write:
\begin{equation}
{\cal C}(t) = t^2 + \frac{t}{2} - \int_0^t\rmd\tau\; B(\tau) \; ,
\quad B(t) = \int_0^t\rmd\tau\; b_2(\tau) \; .
\end{equation}
For a given statistical spectrum $\{E_\alpha\}$ the two-point form factor 
$b_2(t)$ is defined in the equations~(\ref{calC}) and (\ref{defb2}). For the 
GOE and the GUE case, the two-point form factors are well known~\cite{Meh91}, 
and for the Poisson case, $b_2(t)$ is simply zero. Due to the lack of an
appropriate reference, we shall calculate $b_2(t)$ for the picket-fence case.

\subsection{\label{APIF} Picket-fence}

In the picket-fence case, we may assume, that $E_\alpha = \alpha$. 
With the definition 
$\xi(t)= \sum_{\gamma =1}^N \exp(2\pi\rmi(\oE_\gamma-\oE_\alpha)\, t)$, we 
may write:
\begin{equation}
\xi(t) = N^{-1}\sum_{\alpha=1}^N q^\alpha \sum_{\beta=1}^N q^{-\beta} \; ,
\quad q= \rme^{2\pi\rmi\, t} \; ,
\end{equation}
which allows to obtain $\xi(t)$ in closed form. Namely, using the relation 
$\sum_{\alpha=0}^n q^\alpha = (1- q^{n+1})/(1-q)$, we get:
\begin{equation}
\fl \xi(t)= N^{-1}\left( \frac{1- q^{N+1}}{1-q} - 1\right)\left(
   \frac{1- q^{-N-1}}{1-q^{-1}} - 1\right)
= N^{-1}\; \frac{(1-q^N)(1-q^{-N})}{(1-q)(1-q^{-1})} \; ,
\end{equation}
so that
\begin{equation}
\fl \xi(t)= N^{-1}\; \frac{2- q^N - q^{-N}}{2- q - q^{-1}}
 = N^{-1}\; \frac{1- \cos(2\pi\, Nt)}{1- \cos(2\pi\, t)}
 = N^{-1}\; \frac{\sin^2(\pi\, Nt)}{\sin^2(\pi\, t)} \; .
\end{equation}
Now, we will show that $\xi(t)\to \sum_{n\in\mathbb{Z}}\delta(t-n)$ in the 
limit $N\to\infty$. As $\xi(t)$ is periodic with period one, it is
sufficient to consider $t\in (-\case{1}{2},\case{1}{2})$. Let $f(t)$ be an 
arbitrary smooth function, with support in $(-\case{1}{2},\case{1}{2})$. Then:
\begin{eqnarray}
\fl \int\rmd t\; f(t)\; \xi(t) \approx N^{-1}\int_{-N^{-1}}^{N^{-1}}\rmd t\; 
   f(t)\; \frac{\sin^2(\pi\, Nt)}{\sin^2(\pi\, t)} 
\approx \frac{f(0)}{N}\int_{-N^{-1}}^{N^{-1}}\rmd t\; \frac{\sin^2(\pi\, Nt)}
   {\pi^2\, t^2} \nonumber\\
\lo\approx N\; f(0)\int_{-N^{-1}}^{N^{-1}}\rmd t\; \cos^2(\pi\, Nt/2)
= \frac{2}{\pi}\; f(0)\int_{-\pi/2}^{\pi/2}\rmd s\; \cos^2(s) 
\approx f(0) \; .
\end{eqnarray}
This proves the supposition, so that we can conclude:
\begin{equation}
 \sum_{n\in\mathbb{Z}}\delta(t-n) = 1 + \delta(t) - b_2(t) 
\quad\Rightarrow\quad
b_2(t)= 1 - \sum_{n\ne 0} \delta(t-n) \; .
\end{equation}

\subsubsection*{The first and second integral of $b_2(t)$}

The first integral of the two-point form factor is:
\begin{equation}
B(t)= \int_0^t\rmd\tau\left[ 1- \sum_{n\ne 0} \delta(t-n)\right] 
    = t - [t] \qquad [t] = \max_{n\in\mathbb{N}} (n\le t) \; ,
\end{equation}
while for the second integral, we obtain:
\begin{equation}
\int_0^t\rmd\tau\; B(\tau) = \frac{t^2}{2} 
                    - [t] \left( t - \frac{[t]+1}{2}\right) \; .
\end{equation}

\subsection{\label{AGOE} GOE case}

The two-point form factor for the GOE case reads \cite{Meh91}:
\begin{eqnarray}
\fl b_2(t) = t\; \ln(2t+1) - 
   \cases{ 2t - 1             &: $0<t<1$\\
           1 + t \; \ln(2t-1) &: $1<t$}    \nonumber\\
= 1 -2t +  t\; \ln(2t+1) + \theta(t-1)\left[ 2(t-1) - t\; \ln(2t-1) 
   \right] \; ,
\end{eqnarray}
where $\theta(t)$ is the unit step function. Its first integral gives:
\begin{eqnarray}
\fl B(t) = \int_0^t\rmd\tau\; b_2(\tau) 
      = t - t^2 + \int_0^t\rmd\tau\; \tau\; \ln(2\tau+1) \nonumber\\
+ \theta(t-1) \left[ \int_0^{t-1}\rmd\tau\; 2\tau 
            - \int_1^t\rmd\tau\; \tau\; \ln(2\tau -1) \right] \nonumber\\
\lo= \frac{5(t-t^2)}{4} + \frac{t^2-1/4}{2}\; \ln(2t+1) \nonumber\\
+ \theta(t-1) \left[ (t-1)^2 + \frac{(t-1)(t+2)}{4} 
            - \frac{t^2-1/4}{2}\; \ln(2t-1) \right] \; ,
\end{eqnarray}
while the second integration yields:
\begin{eqnarray}
\fl \int_0^t\rmd\tau\; B(\tau) = \frac{5(t^2/2-t^3/3)}{4} 
   + \frac{1}{2}\int_0^t\rmd\tau\; (t^2-1/4)\; \ln(2\tau+1) + \theta(t-1)
\nonumber\\
\times \left[ \frac{5(t-1)^3}{12} + \frac{3(t-1)^2}{8}
   - \frac{1}{2}\int_1^t\rmd\tau\; (t^2-1/4)\; \ln(2\tau-1) \right] \; .
\label{AGOE:BB}\end{eqnarray}
The two remaining integrals give:
\begin{eqnarray}
\fl A_1(t) = \frac{1}{2}\int_0^t\rmd\tau\; (t^2-1/4)\; \ln(2\tau+1)
 = \frac{1}{16}\int_1^{2t+1}\rmd s\; (s-2)s\; \ln s \nonumber\\
\lo= \frac{1}{16}\left[ (y/3-1) y^2\; \ln y - \frac{y^3-1}{9} 
   + \frac{y^2-1}{2} \right]_{y= 2t+1}\\
\fl A_2(t) = \frac{1}{2}\int_1^t\rmd\tau\; (t^2-1/4)\; \ln(2\tau-1)
= \frac{1}{16}\int_1^{2t-1}\rmd s\; (s+2)s\; \ln s \nonumber\\
\lo= \frac{1}{16}\left[ (y/3+1) y^2\; \ln y - \frac{y^3-1}{9} 
   - \frac{y^2-1}{2} \right]_{y= 2t-1} \; ,
\end{eqnarray}
where we have used that:
\begin{equation}
\fl \int_1^y\rmd s\; s\; \ln s = \frac{y^2\; \ln y}{2} - \frac{y^2-1}{4}
\qquad
\int_1^y\rmd s^2\; s\; \ln s = \frac{y^3\; \ln y}{3} - \frac{y^3-1}{9} \; .
\end{equation}

\subsection{\label{AGUE} GUE case}

In this case, the two-point form factor is simply:
\begin{equation}
b_2(t)= \cases{ 1-t &: $0< t<1$\\
                  0 &: $1< t$} \; .
\end{equation}
We give the result for the second integral directly:
\begin{equation}
\int_0^t\rmd\tau\; B(\tau) = \cases{ (1- t/3) t^2/2 &: $0< t< 1$ \\
                                      t/2 - 1/6     &:    $1< t$}   \; .
\end{equation}

\section{\label{A_Pert} Time independent perturbation theory}

Consider the Hamiltonian $H = H_0 + \lambda\; V$. If $H_0$ has a picket-fence
spectrum, we may use the standard (Rayleigh-Schr\" odinger)
perturbation theory to compute low order approximations to the eigenvectors
of $H$. A compact derivation of the perturbation series can be obtained with
the help of the Greens functions for $H$ and $H_0$ (see 
reference~\cite{Messiah79}):
\begin{equation}
\fl G(z) = \frac{1}{z-H} \; , \quad G_0(z) = \frac{1}{z-H_0} 
\quad\Rightarrow\quad G(z) = [1- \lambda\; G_0(z)\; V]^{-1}\; G_0(z) \; .
\end{equation}
Assume that $H_0$ is diagonal in the basis used: 
$\la\alpha|H_0\; \beta\ra = \delta_{\alpha\beta}\; \oE_\alpha$, and define 
$|e_\alpha\ra$ as the eigenvector of $H$, such that:
\begin{equation}
\la e_\alpha|H\; e_\beta\ra = \delta_{\alpha\beta}\; E_\alpha \qquad 
|e_\alpha\ra \to |\alpha\ra \qquad E_\alpha \to \oE_\alpha
\end{equation}
as $\lambda\to 0$. Then we may write the projector onto the eigenstate 
$|e_\alpha\ra$ as follows:
\begin{equation}
P_\alpha = |e_\alpha\ra\; \la e_\alpha| 
         = \frac{1}{2\pi\rmi} \oint\rmd z\; G(z) \; ,
\end{equation}
where the integration path is a simple loop enclosing exclusively the 
pole $\oE_\alpha$ of $G_0(z)$ and the pole $E_\alpha$ of $G(z)$. Then we may 
obtain the perturbation series as follows:
\begin{equation}
\fl P_\alpha^{(n)} = |e_\alpha^{(n)}\ra\; \la e_\alpha^{(n)}| 
 = \sum_{k=0}^n \lambda^k\; \tilde T_\alpha^{(k)} \; , \quad 
\tilde T_\alpha^{(k)} = \frac{1}{2\pi\rmi} \oint\rmd z\; [G_0(z)\; V]^k\; 
     G_0(z) \; .
\label{AP_defP}\end{equation}
The $\tilde T_\alpha^{(k)}$ are operators, and the tilde is used to distinguish
them from the matrices $T^{(k)}$ introduced below. Note that so far, the 
projectors $P_\alpha^{(n)}$ are not normalised, so that an approximation to 
the absolute value squared of a matrix element of $O$ reads:
\begin{equation}
\fl O_{\xi\alpha}^2 = \frac{\la\xi|P_\alpha\; \xi\ra}
                       {\sum_\chi \la\chi|P_\alpha\; \chi\ra}
 = \frac{\delta_{\xi\alpha} + \sum_{k=1}^n \lambda^k\; 
     \la\xi|\tilde T_\alpha^{(k)}\; \xi\ra}{1 + \sum_{l=1}^n \lambda^l\; 
     \sum_{\chi\ne\alpha} \la\chi|\tilde T_\alpha^{(l)}\; \chi\ra}
= \frac{\delta_{\xi\alpha} + \sum_{k=1}^n \lambda^k\; T_{\xi\alpha}^{(k)}}
     {1 + \sum_{l=1}^n \lambda^l\; S_\alpha^{(l)}} \; ,
\end{equation}
where $S_\alpha^{(l)} = \sum_{\chi\ne\alpha} T_{\chi\alpha}^{(l)}$.
As a final step, one may expand this expression in a Taylor series in 
$\lambda$ up to any desired order.

In what follows, we will calculate the matrix elements of $O$ up to forth
order in $\lambda$ and then average the results over the perturbation $V$ which
is assumed to be a GOE matrix. To this end, we first consider
the operators $\tilde T_\alpha^k$, defined in~(\ref{AP_defP}):
\begin{eqnarray}
\fl \tilde T_\alpha^{(0)} = \frac{1}{2\pi\rmi}\oint\rmd z\; G_0(z) 
= |\alpha\ra \; \la \alpha| = p_\alpha \nonumber\\
\fl \tilde T_\alpha^{(1)} = \frac{1}{2\pi\rmi}\oint\rmd z\; G_0(z) \; V\; 
   G_0(z)
 = p_\alpha\; V\; G'_0 + G'_0\; V p_\alpha \; , \quad 
G'_0 = \sum_{\beta\ne\alpha} |\beta\ra \; \frac{1}{\alpha-\beta} \; \la\beta|
\nonumber\\
\fl \tilde T_\alpha^{(2)} = \frac{1}{2\pi\rmi}\oint\rmd z\; 
     G_0(z) \; V\; G_0(z) \; V\; G_0(z) \nonumber\\
\lo= p_\alpha\; V\; G'_0\; V\; G'_0 + G'_0\; V\; p_\alpha\; V\; G'_0
     + G'_0\; V\; G'_0 \; V\; p_\alpha \nonumber\\
\fl \quad\vdots \nonumber\\
\fl \tilde T_\alpha^{(n)} = \sum_{k=0}^n [G'_0\; V]^{k}\; p_\alpha\;
     [V\; G'_0]^{n-k} \; .
\end{eqnarray}
For the matrices $T^{(k)}$, these formulae imply that: $T^{(0)} = 1$, 
$T^{(1)} = 0$ and $T^{(k)}$ has only zeros on the diagonal for all $k>1$.

Our aim is it to calculate the average of products of the form 
$\la O^2_{\xi\alpha}\, O^2_{\mu\alpha}\ra_V$. These are in fact just the 
one-vector averages of order four of this particular ensemble of 
(approximately) orthogonal matrices. To do so we first expand the squared
matrix elements of $O$ in terms of the perturbation parameter $\lambda$. Up
to fourth order, we get ($\xi\ne\alpha$):
\begin{eqnarray}
\fl O^2_{\alpha\alpha} &= 1 - \sum_{l=2}^4 \lambda^l\; S_\alpha^{(l)}
     + \left( \lambda^2\; S_\alpha^{(2)}\right)^2
= 1 - \lambda^2\; S_\alpha^{(2)} - \lambda^3\; S_\alpha^{(3)}
     - \lambda^4\; \left( S_\alpha^{(4)} - {S_\alpha^{(2)}}^2 \right) \\
\fl O^2_{\xi\alpha} &= \sum_{k=2}^4 \lambda^k\; 
     T_{\xi\alpha}^{(k)}\; \left( 1 - \lambda^2\; 
     S_\alpha^{(2)} \right)
= \lambda^2\; T_{\xi\alpha}^{(2)} + \lambda^3\; 
     T_{\xi\alpha}^{(3)} + \lambda^4 \left( T_{\xi\alpha}^{(4)} - 
     T_{\xi\alpha}^{(2)}\; S_\alpha^{(2)} \right) \; .
\end{eqnarray}
Now, the desired one-vector averages are easily constructed. To this end we
assume that the unperturbed energies form the picket-fence spectrum: 
$\oE_\alpha = \alpha$, and we consider the limit $N\to\infty$. Thus, it 
remains to perform the average over the GOE-matrix $V$. Henceforth, we use 
simple angular brackets without subscript to denote such an average. For
$\alpha\ne\xi\ne\mu\ne\alpha$, we obtain:
\begin{eqnarray}
\fl \la O_{\alpha\alpha}^4\ra = 1 - 2\lambda^2\; \la S_\alpha^{(2)}\ra
     - 2\lambda^3\; \la S_\alpha^{(3)}\ra - 2\lambda^4\left(
     \la S_\alpha^{(4)}\ra - \lla {S_\alpha^{(2)}}^2\rra \right)
     + \lambda^4\; {\la S_\alpha^{(2)}}^2\ra \nonumber\\
\lo= 1 - 2\lambda^2\; \la S_\alpha^{(2)}\ra - \lambda^4\left(
     2\; \la S_\alpha^{(4)}\ra - 3\; \lla {S_\alpha^{(2)}}^2\rra \right) \\
\fl \la O_{\xi\alpha}^4\ra = \lambda^4\; \lla {T_{\xi\alpha}^{(2)}}^2\rra \\
\fl \la O_{\alpha\alpha}^2\; O_{\xi\alpha}^2\ra = \lla \left(1 - \lambda^2\;
     S_\alpha^{(2)}\right) \left( \lambda^2\; T_{\xi\alpha}^{(2)} + 
     \lambda^3\; T_{\xi\alpha}^{(3)} + \lambda^4 \left[ T_{\xi\alpha}^{(4)}
   - T_{\xi\alpha}^{(2)}\; S_\alpha^{(2)} \right]
     \right) \rra \nonumber\\
\lo= \lambda^2\; \la T_{\xi\alpha}^{(2)}\ra + \lambda^4\left( 
     \la T_{\xi\alpha}^{(4)}\ra - 2\; 
     \la T_{\xi\alpha}^{(2)}\; S_\alpha^{(2)}\ra \right) \\
\fl \la O_{\xi\alpha}^2\; O_{\mu\alpha}^2\ra = \lambda^4\; 
     \la T_{\xi\alpha}^{(2)} \; T_{\mu\alpha}^{(2)}\ra \; .
\end{eqnarray}
There are in total eight different quantities to average:
\begin{eqnarray}
\fl \la T_{\xi\alpha}^{(2)}\ra = \lla \frac{1}{\alpha-\xi}\; V_{\xi\alpha}\;
     V_{\alpha\xi}\; \frac{1}{\alpha-\xi}\rra = \frac{1}{(\alpha-\xi)^2} \\
\fl \la S_\alpha^{(2)}\ra = \sum_{\xi\ne\alpha} \la T_{\xi\alpha}^{(2)}\ra
  = \frac{\pi^2}{3} \\
\fl \la {T_{\xi\alpha}^{(2)}}^2\ra = \lla \left(
     \frac{V_{\xi\alpha}^2}{(\alpha-\xi)^2}\right)^2 \rra
 = \frac{3}{(\alpha-\xi)^4} \\
\fl \la T_{\xi\alpha}^{(2)}\; T_{\mu\alpha}^{(2)}\ra = 
   \lla \frac{V_{\xi\alpha}^2}
     {(\alpha-\xi)^2}\; \frac{V_{\mu\alpha}^2}{(\alpha-\mu)^2} \rra
 = \frac{1}{(\alpha-\xi)^2\; (\alpha-\mu)^2} \\
\fl \la {S_\alpha^{(2)}}^2\ra = \sum_{\{\xi\ne\mu\} \ne\alpha}
     \la T_{\xi\alpha}^{(2)}\; T_{\mu\alpha}^{(2)}\ra + \sum_{\xi\ne\alpha}
     \la {T_{\xi\alpha}^{(2)}}^2\ra \nonumber\\
\lo= \sum_{\{\xi\ne\mu\} \ne\alpha}
     \frac{1}{(\alpha-\xi)^2\; (\alpha-\mu)^2} + \sum_{\xi\ne\alpha}
     \frac{3}{(\alpha-\xi)^4} \nonumber\\
\lo= \left(\sum_{\xi\ne\alpha} \frac{1}{(\alpha-\xi)^2}
     \right)^2 + \sum_{\xi\ne\alpha} \frac{2}{(\alpha-\xi)^4}
 = \frac{7\; \pi^4}{45} \\
\fl \la T_{\xi\alpha}^{(4)}\ra = \sum_{\mu,\nu \ne \alpha} \left\{ \lla
     \frac{1}{\alpha-\xi}\; V_{\xi\alpha}\; V_{\alpha\mu}\;
     \frac{1}{\alpha-\mu}\; V_{\mu\nu}\; \frac{1}{\alpha-\nu}\; V_{\nu\xi}\;
     \frac{1}{\alpha-\xi}\rra \right. \nonumber\\
+ \lla \frac{1}{\alpha-\xi}\; V_{\xi\mu}\;
     \frac{1}{\alpha-\mu}\; V_{\mu\alpha}\; V_{\alpha\nu}\;
     \frac{1}{\alpha-\nu}\; V_{\nu\xi}\; \frac{1}{\alpha-\xi}\rra \nonumber\\
 \left. + \lla
     \frac{1}{\alpha-\xi}\; V_{\xi\mu}\; \frac{1}{\alpha-\mu}\; V_{\mu\nu}\;
     \frac{1}{\alpha-\nu}\; V_{\nu\alpha}\; V_{\alpha\xi}\; 
     \frac{1}{\alpha-\xi}\rra \right\} \nonumber\\
\lo= \frac{1}{(\alpha-\xi)^2}\left\{ \sum_{\nu\ne\alpha} 
   \frac{\la V_{\xi\alpha}^2\; V_{\xi\nu}^2\ra}
   {(\alpha-\xi)\; (\alpha-\nu)} + \sum_{\mu\ne\alpha} \frac{
   \la V_{\xi\mu}^2\; V_{\mu\alpha}^2\ra}{(\alpha-\mu)^2} + \sum_{\mu\ne\alpha}
   \frac{\la V_{\xi\mu}^2\; V_{\xi\alpha}^2\ra}{(\alpha-\mu)\; (\alpha-\xi)} 
     \right\} \nonumber\\
\lo= \frac{1}{(\alpha-\xi)^4} + \frac{\pi^2}{3}\; \frac{1}{(\alpha-\xi)^2} \\
\fl \la S_\alpha^{(4)}\ra = \sum_{\xi\ne\alpha} \la T_{\xi\alpha}^{(4)}\ra
 = \frac{\pi^4}{45} + \frac{\pi^4}{9} = \frac{2\; \pi^4}{15} \\
\fl \la S_\alpha^{(2)}\; T_{\xi\alpha}^{(2)}\ra = 
   \la {T_{\xi\alpha}^{(2)}}^2\ra 
 + \sum_{\mu\ne\xi,\alpha} \la T_{\xi\alpha}^{(2)}\; T_{\mu\alpha}^{(2)}\ra 
 = \frac{3}{(\alpha-\xi)^4} + \sum_{\mu\ne\xi,\alpha} 
     \frac{1}{(\alpha-\xi)^2\; (\alpha-\mu)^2} \nonumber\\
\lo= \frac{2}{(\alpha-\xi)^4} + \frac{\pi^2}{3}\; \frac{1}{(\alpha-\xi)^2} \; .
\end{eqnarray}

\end{appendix}

\section*{References}

\bibliographystyle{unsrt}
\bibliography{/home/gorin/General/Bib/amol,/home/gorin/General/Bib/deco,/home/gorin/General/Bib/ranh,/home/gorin/General/Bib/semic,/home/gorin/General/Bib/stas,/home/gorin/General/Bib/books,/home/gorin/General/Bib/preprint}

\end{document}